\documentclass{aa}
\usepackage{epsfig}
\def \eg {{\it e.g.} }
\def \etal {{\it et al.} }
\def \ie {{\it i.e. } }
\def \tomega {\tilde\omega}
\def \dss {\displaystyle}
\newcommand{\new}[1]{{{#1}}}

\begin{document}

\title{Excitation of warps by spiral waves in galaxies by non-linear coupling}

\author{F.~Masset \and M.~Tagger}

\offprints{F.~Masset}

\institute{CEA, DSM, DAPNIA, Service d'Astrophysique, CE-Saclay, 91191 Gif-sur-Yvette Cedex, France}

\date{Received 1 September 1995 / Accepted 25 January 1996}

\thesaurus{318, 747--767 (1997)}
\maketitle

\begin{abstract}

We present a new mechanism to explain the frequently observed and thus 
certainly
permanent warping of spiral galaxies. We consider the \new{possibility of 
non-linear} coupling between the
spiral wave of the galaxy and two warp waves\new{, such that the former, 
which is linearly unstable and extracts energy and angular momentum from 
the inner regions of the galactic disk, can continuously feed the latter}. 
We derive an expression for  
the
coupling coefficient in the WKB approximation. We show that the coupling is
too weak in the stellar disk, except at the Outer Lindblad Resonance where 
the
spiral slows down and is efficiently coupled to warp waves. There, the 
spiral
can be almost completely converted into ``transmitted'' warps, which we 
can observe
in HI, and a ``reflected'' one, which we can observe as a corrugation. Our 
mechanism
reproduces the observed amplitudes of the warp and of the corrugation, 
and might
explain related phenomena such as the behavior of the line of nodes
of the warp. Furthermore we show that the energy and momentum fluxes of 
observed 
spirals and warps are of the same order of magnitude, adding a strong 
point in favor of this model.

\keywords{Galaxies : warping, spiral waves, Lindblad resonances -- Nonlinear coupling}

\end{abstract}

\section{Introduction}

Spiral galaxies often show a warping of their external parts, observed in HI. Such
warps appear like an ``S'' or an integral sign for galaxies observed edge-on (Sancisi,
1976).

This phenomenon, known since 1957, has led to theoretical difficulties. 
Warps are bending waves, and we correctly understand their propagation
(see Hunter, 1969a for a dispersion relation in an infinitely thin disk,
and see Nelson 1976a and 1976b, Papaloizou and Lin 
1995, Masset and Tagger 1995 for the dispersion
relation taking into account finite thickness and compressional effects), 
but we do not know the mechanism
responsible for their excitation. It cannot
be systematically justified by tidal effects (Hunter and Toomre 1969),
since we observe warping of very
isolated spiral galaxies. It cannot be explained by a temporary excitation,
since a warp propagates radially in the disk is not reflected at its 
edge, so that it should disappear over a few galactic years (Hunter, 1969b). 
Attempts have been made to connect
the existence of warps and the properties of haloes (Sparke, 1984a, Sparke and
Casertano 1988,
Hofner and Sparke 1994). The mechanism proposed by Sparke
and Casertano (1988)
implies an {\it ad hoc} misalignment angle between the normal axis of the galactic
plane and the ``polar'' axis of the halo. This mechanism has recently encountered
self-consistency difficulties (Dubinski and Kuijken 1995). Furthermore, it relies on the
dubious hypothesis of a rigid and unresponsive halo.
Several explanations have also been proposed, such as the 
infall of primordial matter (Binney, 1992), 
the effect of the dynamical pressure of the intergalactic medium,
etc\ldots.  For a review of all the proposed mechanisms, see Binney (1992).

Binney (1978, 1981) has considered the possibility 
of a resonant coupling between the vertical
motion of a star and the variation of the galactocentric force, due to
a halo or a bar. He concluded that a bar could be responsible for the observed warps
and corrugations.
Sparke (1984b) has also explored this possibility, and considered the growth of a warp
from a bar or a triaxial halo. She found that a bar was unlikely to be responsible
for a warp, but she emphasized that a triaxial halo could quite well reproduce observed
warps.

We consider here another mechanism, the non-linear coupling between a 
spiral wave and two warp waves. Non-linear coupling between spirals and 
bars has already been found (Tagger \etal 1987 and Sygnet \etal 1988) to 
provide a convincing explanation for certain behaviors observed in 
numerical simulations 
(Sellwood 1985), or observed in Fourier Transforms
of pictures of face-on galaxies. The relative amplitude of spiral or warp 
waves (the ratio of the perturbed potentials 
to the axisymmetric one) is about $\sim 0.1$ to $\sim 
0.3$ (see Strom \etal 1976). Non-linear coupling involves terms of second 
order in the perturbed potential, while linear propagation is described 
by first order terms.  Classically one would thus believe that non-linear 
coupling is weak at such small relative amplitudes. However in the 
above-mentioned works it was found that the presence of resonances could 
make the coupling much more efficient if the wave frequencies are such 
that their resonances (\ie the corotation of one wave and a Lindblad 
resonance of another one) coincide. At this radius the non-linear terms 
become comparable with the linear ones, so that the waves can very 
efficiently exchange energy and angular momentum. Indeed in Sellwood's 
(1985) simulations, as discussed by Tagger \etal (1987) and Sygnet \etal 
(1988), an ``inner'' spiral or bar wave, as it reaches its corotation 
radius, transfers the energy and angular momentum extracted from the 
inner parts of the disk to an ``outer'' one whose ILR lies at the same 
radius, and which will transfer them 
further out, and ultimately deposit them at its OLR. In this process the 
energy and momentum are thus transferred much farther radially than they would 
have been by a single wave, limited in its radial extent by the peaked 
rotation profile.

We will show here that a similar mechanism, now involving one spiral and 
two warp waves is not only possible (by the ``selection rules'' 
associated with their parity and wavenumbers), but also very efficient if 
the same coincidence of resonances occurs. This allows the spiral wave, 
as it reaches its OLR (and from linear theory deposits the energy and 
momentum extracted from the inner regions of the disk) to transfer them 
to the warps which will carry them further out. 

Unlike Tagger \etal (1987) and Sygnet \etal 
(1988), we will throughout this paper restrict our analysis to gaseous 
rather than stellar disks, described from hydrodynamics rather than from 
the Vlasov equation. \new{The reason is that our interest here lies 
mainly in the excitation of the warps, which propagate essentially in the 
gas (indeed the outer warp is observed in HI, and the corrugation is most 
likely 
(Florido \etal 
1991) due to
the motion of the gaseous component of the galactic disk).
On the other hand, the spiral wave propagates in the stellar as well 
as the gaseous disk. The difference is important only in the immediate 
vicinity of Lindblad resonances, where the spiral wave is absorbed; as a 
consequence, 
its group velocity vanishes at the resonances. Since the group velocity 
of the waves will appear as an important parameter, we will choose to keep 
the analytic coupling coefficient derived from the
hydrodynamic analysis, but we will introduce, for the spiral 
density wave, the
group velocity of a stellar spiral. From the physics involved this will 
appear as a reasonable approximation; furthermore it should only 
underestimate the coupling efficiency, since it does not include the 
resonant stellar motions near the resonance. 

On the other hand, we will show that non-linear coupling is efficient 
only in a narrow annulus close to the OLR of the spiral, over a scale length 
similar to the one of Landau damping. We will thus conclude that the two 
processes are in direct competition, with the spiral transferring its 
energy and momentum, in part to the stars by Landau damping, and in part 
to the warps which will transfer them further outward, the exact 
repartition between these mechanisms presumably depending on detailed 
characteristics of the galactic disk.}  

The paper is organized as follows: 
in a first part we will introduce the notations, and the selection rules 
relative to the coupling. In a second part, we will derive the coupling
coefficient from the hydrodynamic equations expanded to second order in 
the perturbed quantities, 
and we will try and simplify it. In a third part, we will analyze the
efficiency of the coupling, together with the locations where it may 
occur. In the last sections we will compare
our predictions to the observations, and we will propose some possible
observational tests of our mechanism.

Some of the computations are tedious and lengthy. For the sake of clarity,
they are developed in appendices, so as to retain in the main text only
the principal results and the physical discussions.

\section{Formalism and selection rules}
\subsection{Notations}
We develop all our computations in the well-known {\it shearing sheet}
approximation, which consists in rectifying a narrow annulus around
the corotation radius of the spiral wave into a Cartesian slab. The 
results given by an exact
computation taking into account the cylindrical geometry of the galaxy
would differ by some metric coefficients, but would not differ
physically from the {\it shearing sheet} predictions, so that the main 
conclusions would remain valid. We call $x$ the radial coordinate,
with its origin at the corotation of the spiral, and oriented outward.
We call $y$ the azimuthal coordinate, oriented in the direction of the 
rotation, and $z$ the vertical one, so as to construct
a right-oriented frame $xyz$.

The hydrodynamic quantities are the density $\rho$, the speed components 
$(U,V,W)$, and the gravitational potential $\phi$. We assume that the 
disk is 
isothermal, with a uniform temperature, in order to avoid unnecessary 
complexity. Thus we can write: $P = a^2\rho$, where $P$ is the pressure
and $a$ the isothermal sound speed.

The perturbed quantities are denoted with subscripts $1$, $2$ or $S$ referring to the
wave involved (one of the two warps or the spiral wave), and equilibrium
quantities with a subscript $0$.

The epicyclic frequency is denoted by $\kappa$, the rotation frequency by 
$\Omega$, Oort's first constant by $A=\Omega'r/2$, and we have the relation:
\[\kappa^2 = 4\Omega(\Omega+A)\]

We call $\mu$ the vertical characteristic frequency of the disk.  We have:

\[\mu^2 = 2\Omega^2-\kappa^2\]

(see Hunter and Toomre, 1969).
 This 
frequency plays the same role for bending waves as $\kappa$ does for spiral 
waves; it should not be confused with the frequency $\nu_z$ of the vertical 
motion of individual particles, which is usually much higher. Indeed 
individual particles move vertically in the potential well of the disk, 
giving them the high frequency $\nu_z$. On the other hand, when one 
considers motions of the whole disk, the potential well moves together 
with the disk and exerts no restoring force. This leaves only a weaker 
restoring force, giving the vertical frequency $\mu$ as discussed \eg in 
Hunter and Toomre (1969) or Masset and Tagger (1995).

For each of the three waves we consider, we denote by
$\omega$ its frequency  in the galactocentric frame, $m$ its azimuthal 
wavenumber ($m=2$ for a
two-armed spiral, and $m=1$ for an ``integral-sign'' warp), $\tomega=\omega-m\Omega(r)$ its 
frequency in a local frame rotating with the  matter. In the peculiar
case of the  shearing sheet, we note $k_y = m/r$ where $r$ is the distance
to the galactic center, and $k_y$ the azimuthal wavenumber.

We will have to perform part of our computations in the WKB 
approximation, \ie assume that the radial wavevector varies relatively 
weakly 
over one radial wavelength; in practice, in the shearing sheet, 
this reduces to the classical ``tightly wound'' approximation, $k_y\ll k_x$. 
We will also use $q=\sqrt{k_x^2+k_y^2}$, the modulus of the ``horizontal'' 
wavevector. In the tightly wound approximation one has $q\simeq |k_x|$.

We also introduce $s=\rho/\rho_0$, and various integrated quantities:
\[\sigma = \int_{-\infty}^{+\infty}\rho(z)dz\]
the perturbed surface density,
\[\Sigma = \int_{-\infty}^{+\infty}\rho_0(z)dz\]
the equilibrium surface density, and:
\[Z = \frac{1}{\Sigma}\int_{-\infty}^{+\infty}z\rho(z)dz\]
the mean vertical deviation of a column of matter from the 
midplane, under the influence of a warp.

\begin{figure}
\psfig{file=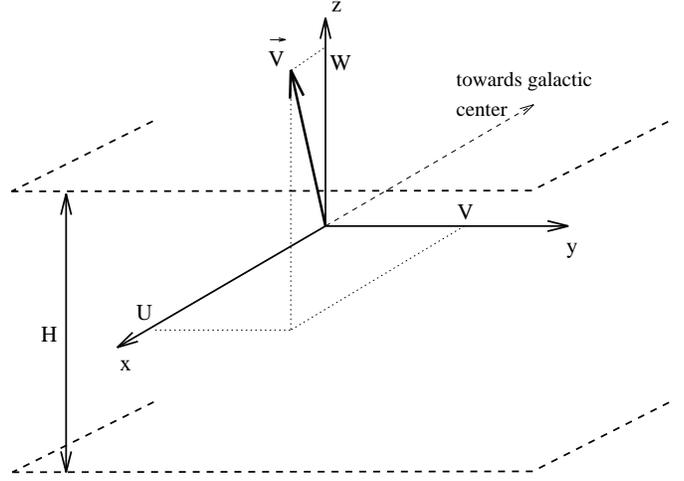,width=\columnwidth}
\caption{\label {fig:not1} \scriptsize
This figure summarizes our main notations relative to perturbed velocity and
disk thickness.}
\end{figure}

We call $H$ the characteristic thickness of the disk. The vertical
density profile in the disk is taken to be {\it consistent}
(see Masset and Tagger, 1995), i.e. it must fulfill simultaneously the
Poisson equation and the hydrostatic equilibrium equation, with the
additional condition that the radial derivatives of the equilibrium potential
do not depend on $z$ throughout the disk thickness: this condition 
results from the hypothesis of a disk which is geometrically thin, $H\ll 
r$, although we do resolve vertically
the perturbed quantities along the vertical direction.

\begin{figure}
\psfig{file=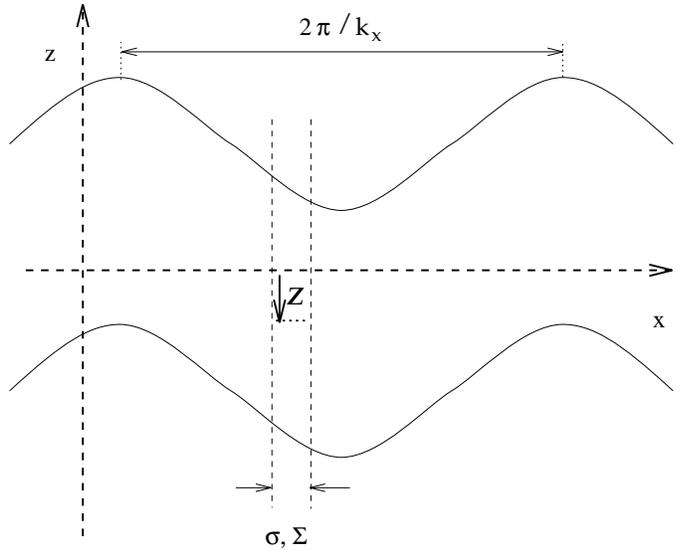,width=\columnwidth}
\caption{\label {fig:not2} \scriptsize
This figure summarizes our notations relative to waves (spiral or warps). The column of
matter has a section unity. At equilibrium (or more generally without spiral wave) it contains
a mass $\Sigma$, and when perturbed by a spiral wave it contains the mass
$\Sigma+\sigma$.}
\end{figure}

We summarize most of our notations in figures \ref{fig:not1} and \ref{fig:not2}
and in table \ref{tab:not1}.

\begin{table}
\begin{center}
\begin{tabular}{|c|l|} 
\hline \hline
{\em Symbol} &
        \multicolumn{1}{c|}{\em corresponding quantity} \\
\hline \hline
$a$             & \parbox{6cm}{\scriptsize Sound speed} \\
\hline
$\Omega$        & \parbox{6cm}{\scriptsize Equilibrium angular velocity of matter} \\
\hline
$A$             & \parbox{6cm}{\scriptsize Oort's first constant. $A =1/2rd\Omega/dr$} \\
\hline
$\kappa$        & \parbox{6cm}{\scriptsize Epicyclic frequency.
$\kappa^2 = 4\Omega(\Omega+A)$} \\
\hline
$\mu$           & \parbox{6cm}{\scriptsize ``Global vertical frequency''. $\mu^2 = 2\Omega^2-\kappa^2$}\\
\hline
$\omega$        & \parbox{6cm}{\scriptsize Wave frequency in galactocentric frame.}\\
\hline
$\tilde\omega$  & \parbox{6cm}{\scriptsize Wave frequency in rotating frame.}\\
\hline
$\phi$          & \parbox{6cm}{\scriptsize Perturbed potential.} \\
\hline
$\rho$          & \parbox{6cm}{\scriptsize Perturbed density.} \\
\hline
$s$             & \parbox{6cm}{\scriptsize Relative perturbed density
($s=\rho/\rho_0$).} \\
\hline
$q$             & \parbox{6cm}{\scriptsize Wave-vector modulus
 ($q=(k_x^2+k_y^2)^{1/2}$)} \\
\hline
\end{tabular}
\end{center}

\caption{\label{tab:not1} \scriptsize
In this table we summarize the main frequencies of our system, and some other
quantities related to warps or spirals.}
\end{table}

\subsection{Notion of coupling and selection rules}
\subsubsection{Mode coupling}
As discussed in the introduction, we consider the coupling between a spiral wave and two
warp waves. This coupling is necessarily a non-linear mechanism. In a linear analysis,
each wave can be studied independently, by projecting perturbed quantities onto
the frequency and the wavevector of each wave. In particular, combining the
hydrodynamic (continuity and Euler) equations and the Poisson equation leads to
the dispersion relation which reads, in the WKB limit:

\[\tomega_S^2 = \kappa^2-2\pi G\Sigma q_S+a^2q_S^2\]
for a spiral and:
\[\tomega_W^2 = \mu^2+2\pi G\Sigma q_W+\frac{\tomega_W^2}{\tomega_W^2-\kappa^2}a^2q_W^2\]
for a warp (see Masset and Tagger 1995, Papaloizou and Lin 1995).

Here we expand the hydrodynamic equations to second order in the perturbed 
quantities. Then the waves do not
evolve independently anymore and, provided that they satisfy selection rules 
which will be discussed below, they can exchange energy and angular 
momentum.

Thus the problem we address can be described as follows:

We consider a spiral wave, with a given flux (its energy density multiplied
by its group velocity), traveling outward, from its corotation to the
Holmberg radius of the galaxy. We will not discuss here how this spiral
has been excited and we do not try to justify its amplitude, but rather 
take it as an observational fact.

As it travels radially the spiral interacts with warps which are always 
present, at all frequencies, at a noise level associated with supernovae 
explosions, remote tidal excitations, etc. 

Our goal is the following: can we find conditions such that, although the 
warps are initially at this low noise level, they can be non-linearly 
coupled to the spiral so efficiently that they absorb a sizable 
fraction of its energy and momentum flux~? And can these conditions be met 
commonly enough to explain the frequent (one might even say general) 
occurrence of warps and their main observational properties~?

We will first discuss the conditions, known as ``selection rules'',  for 
the warps to be coupled to the spiral.
\subsubsection{Selection Rules}
\label{selrule}
The linearized set of equations governing the wave behavior (continuity, 
Euler and Poisson)
is homogeneous and even in $z$, so that any field of perturbations to the 
equilibrium state of the disk can be
considered as composed of two independent parts:
\begin{itemize}
\item{Perturbations whose perturbed density is even in $z$, and thus, due to
hydrodynamic equations, whose perturbed quantities are all even in $z$ except
$W$, which is odd. These perturbations are {\it spiral waves}, since
they imply a perturbed density (a non-vanishing $\sigma$), and a vanishing
$Z$ (so they do not raise the mid-plane of the disk).}
\item{Perturbations whose perturbed density is odd in $z$, and thus whose
perturbed quantities are all odd in $z$, except $W$, which is even.
These perturbations are {\it warps}, since they imply a vanishing 
integrated 
perturbed density and a non-vanishing $Z$, which means that they involve
a global motion of the mid-plane of the disk.}
\end{itemize}

Furthermore, since the coefficients of the equations do not depend on 
time or azimuthal angle, Fourier analysis allows to separate solutions 
identified by their frequency and azimuthal wavenumber. On the other 
hand, since the coefficient do depend on the radius, Fourier analysis in 
$r$ (or $x$ in the shearing sheet) does not allow the definition of a radial 
wavenumber except in the WKB approximation. We will return to 
this below.

So let us consider a spiral wave for which we write:

\[\xi_S \propto e^{i(m_S\theta-\omega_St)}\]

and two warp waves with:
\[\xi_1 \propto  e^{i(m_1\theta-\omega_1t)}\]
and
\[\xi_2 \propto  e^{i(m_2\theta-\omega_2t)}\]

where $\xi$ represents any perturbed quantity. Hydrodynamic 
equations written to second order in perturbed quantities will
contain terms involving the products  $\xi_1\xi_2$,  $\xi_1\xi_2^*$, 
$\xi_1^*\xi_2$, etc.

Let us consider for instance the term $\xi_1\xi_2$ (rules for the other 
products are derived in a similar manner). Its behavior with 
time and azimuthal angle is:
\[\xi_1\xi_2\propto e^{i[(m_1+m_2)\theta-(\omega_1+\omega_2)t]}\]

More physically this product, which appears
from such terms as $V_1.\nabla V_2$ or $\rho_1\nabla \phi_2$, can be 
interpreted as a {\it beat wave}. Fourier analysis in $t$ and $\theta$ 
will give a contribution from this term, at frequency $\omega_1+\omega_2$ 
and wavenumber $m_1+m_2$; thus it can interact with the spiral wave if:
\[\omega_1+\omega_2=\omega_S\]
and
\[m_1+m_2=m_s\]

As discussed above, the equations for the spiral wave are derived from 
the part of the perturbed quantities which is {\it even} in $z$. Our 
third selection rule is thus that the product $\xi_1\xi_2$ be even in $z$.

If we were in a radially homogeneous system, so that waves could also be 
separated by their radial wavenumber, a fourth selection rule would be:
\[k_{x1}+k_{x2}=k_{xS}.\] 
Since this is not the case, we will write a 
coupling coefficient which depends on $x$. This coefficient would vanish by radial 
Fourier analysis if the waves had well-defined radial wavenumbers, unless 
these wavenumbers obeyed this fourth selection rule. We will rather find 
here that the coefficient varies rapidly with $x$ and gives a coupling very 
localized in a narrow radial region. 
\new{This is not a surprise since it also 
occurred in the interpretation by Tagger \etal (1987) and Sygnet \etal 
(1988), in terms of non-linear 
coupling between spiral waves, of the numerical results of Sellwood 
(1985). The localization of the coupling, as will be described later, is 
associated with the presence of the Lindblad resonances. We will find 
that, in practice, it results in an impulsive-like 
generation of the warp waves, in a sense that will be discussed in 
section 4.2.2.

Let us mention here that we have undertaken 
Particle-Mesh numerical simulations in order to confirm this analysis. 
Preliminary 
results in 2D, with initial conditions similar to Sellwood (1985), do show 
the $m=0$ and $m=4$ spiral waves, at the exact 
frequencies and radial location predicted by Tagger \etal (1987) and 
Sygnet \etal 
(1988). The same absence of a selection rule for the radial wavenumber is 
observed. These results will be reported elsewhere, and in a second step 
the simulations will be applied in 3D to the physics described in this paper.

A radial selection rule would allow us to get a very simple expression for the 
efficiency of non-linear coupling - \ie 0 for wave triplets that do not obey it, and 1 
for triplets that do. Here we will have to rely on a more delicate 
integration of the coupling term over the radial extent where it acts. 
This will be done in section 4.2. In particular in 4.2.2, we will show 
that in the vicinity of the OLR of the spiral, even though its WKB 
radial wavenumber is divergent, the main localization of the coupling 
comes from an (integrable) divergence of the coupling coefficient.}

Let us summarize the selection rules:

\begin{itemize}
\item{We first have the condition $m_1+m_2=m_S$. This is obviously 
fulfilled by the most frequent warps (which have $m=1$) and spirals (with 
$m=2$). One should note here that, since we are computing with complex 
numbers but dealing with real quantities, each perturbation is associated 
with its complex conjugate, with wavenumber $-m$ and frequency 
$-\omega^*$. Thus we find a contribution to the warp 1 by the coupling of the 
spiral with the complex conjugate of warp 2, \ie from products of the 
form $\xi_S\xi_2^*$, etc. }
\item{We then have the parity condition, which is obviously fulfilled by 
odd warps and an even spiral.}
\item{Finally the frequency selection rule:
\[\omega_1+\omega_2=\omega_S\]
gives us in principle an infinite choice of pairs of warp waves, since 
the latter can be presumed to form a continuous spectrum. One of our main 
tasks in the present work will be to determine which pair is 
preferentially coupled to the spiral. Our assumption of coupling between 
only three waves will be found valid when we find that actually one such 
pair is strongly favored, so that all the others can be neglected.

The frequency condition can also be written as:
\[\tomega_1+\tomega_2=\omega_1-\Omega+\omega_2-\Omega=\omega_S-2\Omega=\tomega_S\]
\label{partout}
showing that it is also true in the rotating frame at any radius, if it 
is true anywhere.

A second remark concerning this selection rule is that $\omega$ can have 
an imaginary part. 
One is thus restricted to two possible choices : either working in a 
two-time-scales approximation, where the fast scale is that of the 
oscillations ($\omega^{-1}$), while the slow one is both that of  linear growth or damping and of non-linear evolution; or simplify the problem by only 
looking for permanent regimes, keeping $\omega$ real. We will retain the 
second possibility, and in fact consider the non-linear evolution of the 
waves as a function of $x$ as they travel radially, assuming that the 
inner part of the disk feeds the region we consider with spiral waves at 
a constant amplitude. This is sufficient for our main goal, which is to 
show that non-linear coupling is indeed possible and efficient to generate warps. On the 
other hand one should keep in mind that the permanent regime we will find 
may very well be unstable, since it is well known that mode coupling (in 
the classical case of a homogeneous system, much simpler than the one we 
consider) can lead to any type of complex time behavior, \eg limit 
cycles or even strange attractors.}
\end{itemize}

\section{The coupling coefficient}
Once we have the set of selection rules, we can derive an expression of the coupling
coefficient between the spiral and the two warps. In a linear analysis, each
wave propagates independently and is thus subject to a conservation law of the
form:
\[\partial_tE+\frac{1}{r}\partial_r(rc_gE)=0\]
where $\partial_u$ stands for the partial derivative with respect to any  
quantity $u$, $E$ is the energy density of the wave, $c_g$ its group velocity. This relation
simply means that globally the energy of the wave is conserved and
advected at the group velocity $c_g=\partial\tomega/\partial k_x$
(Mark, 1974).
The $r$ and $1/r$ factors in the spatial derivative come from the cylindrical
geometry of the problem. In the framework of the shearing sheet we 
neglect them 
and we may write:

\[d_tE\equiv\partial_tE+\partial_x(c_gE)=0\]

The vanishing right-hand side results from the fact that the wave does 
not exchange energy with other waves or with the particles, and thus is neither amplified
nor absorbed. Mode coupling introduces in the right-hand side a new term 
describing the energy exchange between the waves. Since $E$ is quadratic 
in the perturbed amplitudes, and mode coupling corresponds to going one 
order further in an expansion in the perturbed amplitudes, this new source term will 
be of third order. Its derivation is lengthy and technical, and we give 
it in separate appendices. We summarize it in the next sections.

\subsection{First step : exact expression of the coupling}
The first step consists in deriving an exact expression for this coupling
coefficient. To do this we integrate the hydrodynamic equations over $z$
after transforming their linearized 
parts to write them as 
{\em variational forms}. After a few transformations, we find that the 
total time 
derivative of a variational form, which involves the perturbed quantities 
associated with each 
wave (and which we will interpret as its energy density), is equal to
a sum of terms involving the three waves, which we will interpret
as the coupling part, \ie the energy exchanged between the waves. This 
derivation is done in appendix~\ref{apdx:formel}
for the case of one of the warp waves. Let us just mention
that this first step has the following features:

\begin{itemize}
\item{The only approximations we make are that the disk is isothermal with
a uniform temperature, and that the horizontal velocities in each wave, $U$ and $V$,
are in quadrature, a reasonable assumption
which is asymptotically true in the limit
of the WKB regime.}
\item{We fully resolve the thickness of the disk. Thus, the coupling coefficient
involves integrals over $z$ of products of perturbed quantities related to each wave.}
\item{For the sake of compactness, the derivation is made in tensor formalism. 
This allows us to have a restricted number a terms (in fact eight
terms in the coupling coefficient), although at this
step of the derivation we do not make any hypothesis on the perturbed 
motions associated with a spiral or a warp.}
\end{itemize}

Finally, at the end of this primary step, we obtain for the time evolution
of the energy density of warp 1:

\begin{equation}
\label{eqn:coupstep3}
\frac{d}{dt}\int\rho_0\biggl[(V_1^iV_{1_i}^*+a^2s_1s_1^*)+\biggl(\frac{1}{2}
s_1^*\phi_1+c.c.\biggr)\biggr]
\end{equation}

\[=-\int\rho_0V_{1_i}^*V^j_S\partial_jV^{i*}_2-\int\rho_0V^*_{1_i}V^{j*}_2\partial_jV^i_S\]
\[+\int\rho_0a^2V^*_{1_i}s_S\partial^is^*_2+\int\rho_0a^2V^*_{1_i}s_2^*\partial^is_S\]
\[-\int a^2s_1^*\partial_i(\rho_SV^{i*}_2)-\int a^2s^*_1\partial_i(\rho_2^*V^i_S)\]
\[+\int\phi_1^*\partial_i(\rho_SV_2^{i*})+\int\phi_1^*\partial_i(\rho_2^*V_S^i)+c.c.\]

where $d/dt$ means $\partial/\partial t+\partial_x(c_g.\;\;\;\;)$. The L.H.S.
appears as the total derivative of the energy density of warp~1, 
as expected, and the R.H.S. represents
the coupling term, since each of the integrals involve the perturbed quantities of
the other waves (spiral and warp~2).

\subsection{Second step : analytic result in the WKB approximation}
Equation~(\ref{eqn:coupstep3}) is too complex to be used directly. We 
show in appendix \ref{apdx:vecppe} how it can be simplified by expanding the implicit sums, and explicitly writing the perturbed quantities
associated with the warps or the spiral, and then by using in the 
evaluation of this term the eigenvectors (\ie the various components of 
the perturbation) derived from the linear analysis. We make some assumptions:
\begin{itemize}
\item{We assume that the spiral involves no vertical motions, i.e. $W_S \equiv 0$, and that
horizontal motions do not depend on $z$, i.e. $\partial_zU_S=\partial_zV_S=0$.
These two assumptions lead to an eigenvector which is the one of the infinitely thin
disk approximation, although we vertically resolve the disk.}
\item{Symmetrically, we assume that the vertical speed in a warp is independent of $z$,
\ie: $\partial_zW_1 = \partial_zW_2 = 0$, so that the perturbed density 
is obtained by a simple vertical translation of the equilibrium density 
profile. This is consistent only if there is no horizontal motions. On 
the other hand, the compressibility of the gas disk does introduce 
horizontal motions in the warp wave (Masset and Tagger, 1995); thus, in order  
not to loose a possible
coupling through these horizontal motions, we retain in the expression of 
the coupling coefficient the horizontal velocities due to the warps. This 
may be important near the Lindblad resonances of the warp,
where horizontal motions can become dominant.}
\end{itemize}
At the end of this second step, we get a new expression for $dE_1/dt$
(where $E_1$ is the energy density of warp 1) where we have performed 
the integrals over $z$, and which can be written:

\begin{equation}
\label{eqn:modif1}
\frac{d}{dt}E_1 = \lambda_1 Z_1Z_2\sigma
\end{equation}

where $\lambda_1$ is a long expression which does 
not need to be reproduced 
here, and which depends on $x$. Equation~(\ref{eqn:modif1}) shows that  
warp 1 is coupled to the product of the amplitude of both warps
($Z_1$ and $Z_2$) and to the amplitude of the spiral ($\sigma$). In 
particular, we
see that if the warp does not exist at all in the beginning (i.e. it has a 
vanishing
amplitude), it will never grow. Thus it has to pre-exist at some noise 
level
in order to be allowed to couple with another warp and the spiral. We see 
here an important difference with the harmonic generation 
by a single wave, since in that case the harmonic will be generated even 
it it does not pre-exist. Here we are 
rather
concerned with the production of warps as ``sub-harmonics'' of the spiral. 
They cannot be spontaneously created, as this would correspond to a 
breaking
of azimuthal symmetry.

Mathematically, we see that the L.H.S. of the expression above is second order
with respect to the perturbations amplitudes, and that the R.H.S. is third order.
In fact, the L.H.S. comes from the linear part of hydrodynamic equations, and has
been transformed into a second order expression (a variational
form), and the R.H.S., which comes from the non-linear part of the equations,
has simultaneously been converted into a third order expression. 

In the same manner we can derive, by swapping indices 1 and 2, an expression for
the time evolution of the energy density of warp 2:

\[\frac{d}{dt}E_2 = \lambda_2 Z_1Z_2\sigma\]

In order to close our set of equations, we also have to follow  the behavior of
the spiral.
In the absence of any coupling to other waves, for reasons of global conservation of
energy, the time evolution equation of the spiral is given by:

\[\frac{d}{dt}E_S = -(\lambda_1+\lambda_2) Z_1Z_2\sigma\]

We choose to neglect coupling to other waves here since  they either belong to 
the dynamics of the spiral itself (\eg generation of $m=4$ harmonics), and 
are irrelevant here since we take the spiral as an observational fact, or 
involve other warps: but, as mentioned above, we 
will find below that one pair of warps is preferentially driven by the spiral.

\subsection{Final step : the simplification}
The third and final step is to give an estimate of the coupling coefficients $\lambda$. 
This is done in appendix~\ref{apdx:simpli}. Simplifying this
coefficient implies much discussion on the physics involved, \eg the behavior
of resonant terms near the Lindblad resonances, or the order of magnitude of ratios
of characteristic frequencies for a realistic galactic disk, etc. These 
discussions are fully developed in appendix~\ref{apdx:simpli}, 
and lead us to the result that the coupling
coefficient, anywhere in the disk, is always of the order of 
$\nu_z^2\tomega_S$, \ie:

\begin{equation}
\label{eqn:principal_equation}
\frac{d}{dt}E_1 \sim \nu_z^2\tomega_S Z_1Z_2\sigma
\end{equation}

where $\nu_z$ represents the frequency of vertical oscillations of a test particle
in the rest potential of the disk (it must not be confused with $\mu$, which is
the frequency of global oscillations of the galactic plane, achieved when
the whole disk -- stars and gas-- is coherently moved up and down).
We see that this coefficient does not contain any dependency on 1 and 2, 
so that 
$\lambda_1\sim\lambda_2\sim\nu_z^2\tomega$.

An important difference occurs here with the case of coupling between 
spirals or bars, analyzed by Tagger \etal (1987) and Sygnet \etal (1988). 
In that case the coupling coefficients, obtained from kinetic theory, 
were found to involve two resonant denominators, corresponding to the 
resonances of two of the waves; this made the coupling very efficient if 
the two denominators vanished at the same radius. We do not find such 
denominators here, because the warp vertical velocity (represented by $Z$ 
in equation \ref{eqn:principal_equation}) does not diverge at the Lindblad 
resonance, while the horizontal velocities associated with the spiral do. 
On the other hand we will recover here a similar property of strong 
coupling close to the resonances, because the group velocity of the waves 
becomes small (it goes to zero in the asymptotic limit), so that the 
waves can non-linearly interact for a long time. This will be discussed 
in more details below.

\section{The coupling efficiency}
We have to solve, or at least analyze, the behavior of a system of coupled 
differential equations which is non-linear (since the right-hand-sides
are products of the unknowns). Such a system can have  subtle and varied
solutions, such as limit cycles or strange attractors (chaotic behavior). 
This goes far beyond the scope of this paper and we will restrict 
ourselves to looking for stationary solutions, which might prove to be 
unstable but still will inform us on the efficiency of the coupling 
mechanism. Thus we suppress the partial $t$ derivatives in the
total derivative $d/dt$, making solution much simpler.

\subsection{Search for a stationary solution}
We can write the system of three coupled equations in a slightly 
different manner, by
introducing the energy fluxes ($H_i=c_{g_i}E_i$) of the waves, and
by relating density energies and amplitudes through constants $K_i$ 
derived in appendix \ref{apdx:energy}:

\renewcommand{\arraystretch}{1.5}
\[ \left\{ \begin{array}{l}
E_i = \Sigma K_i^2Z_i^2 \mbox{ for $i=1,2$} \\
E_S = K_S^2\frac{\sigma^2}{\Sigma}
\end{array}
\right.
\]
\renewcommand{\arraystretch}{1}

(see for the derivation of $K_i$);
the energy densities are proportional to the square of the 
amplitudes,
while $K_S$ of the order of magnitude of the sound speed, as expected for 
an
acoustic wave, and $K_1$ and $K_2$ are of the order of the ``spring 
constant'' $\tomega_{(1,2)}^2$).

With these new notations we get:

\renewcommand{\arraystretch}{2.5}

\[ \mbox{(I)} \left\{ \begin{array}{l}
\partial_x(H_1^{1/2}) = \dss \frac{\tomega_s\nu_z^2}{2\Sigma^{1/2}}
\frac{1}{
K_1K_2K_S
}\frac{H_2^{1/2}H_S^{1/2}}{\sqrt{c_{g_1}c_{g_2}c_{g_S}}}
 \\
\partial_x(H_2^{1/2}) = \dss \frac{\tomega_s\nu_z^2}{2\Sigma^{1/2}}
\frac{1}{
K_1K_2K_S 
}\frac{H_1^{1/2}H_S^{1/2}}{\sqrt{c_{g_1}c_{g_2}c_{g_S}}}
 \\
\partial_x(H_S^{1/2}) = \dss -\frac{\tomega_s\nu_z^2}{\Sigma^{1/2}}
\frac{1}{
K_1K_2K_S
}\frac{H_1^{1/2}H_2^{1/2}}{\sqrt{c_{g_1}c_{g_2}c_{g_S}}}
\end{array}
\right.
\]

\renewcommand{\arraystretch}{1}

In order to be able to estimate the behavior of the solutions of system (I), we
must analyze the geometry and localization of the coupling.

\subsection{Localization of the coupling}

\begin{figure}
\psfig{file=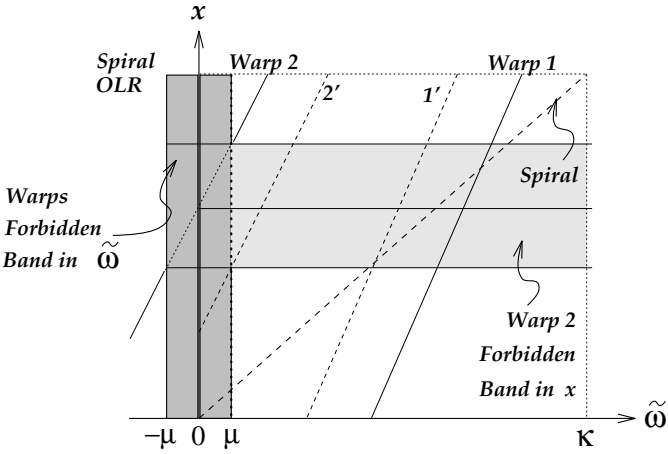,width=\columnwidth}
\caption{\label{fig:diag} \scriptsize
On this figure are presented two pairs of warps which can be coupled with the spiral.
At each $x$, we see on the graph that $\tomega_1+\tomega_2=\tomega_S$, and also that
$\tomega_{1'}+\tomega_{2'}=\tomega_S$. The grey band at the left arises from the
dispersion relation of warps, which cannot propagate where $|\tomega| < \mu$.}
\end{figure}

To this point we have considered the coupling between three waves: the spiral wave, and two
low amplitude warp waves which satisfy the selection rules. But any pair 
of $m=1$ warps obeying the frequency selection rule will also obey the 
other ones, leaving us potentially with a continuous set of warp pairs 
coupled to the spiral. The question is now to determine whether
one or several couples of warps can be preferentially amplified by the spiral wave,
and if so, which one(s) and where. In order to clarify the situation, we present
the problem graphically on figure \ref{fig:diag}. On this figure, each wave is
presented by a curve (a segment in the  shearing sheet approximation, but this
does not lead to a loss of generality). For each curve, $x$ is given by the implicit
relation $\tomega = \omega - m\Omega(x)$, where we have taken $m_S =2$ and
$m_1=m_2=1$. It is an easy matter to check that the slope of the curves 
describing the warps, at a given
$x$, must be twice the slope of the curve describing the spiral. For the reason
already explained in section \ref{partout}, if two warps satisfy the $\tomega$-selection
rule at some $x$, they satisfy it at any $x$. Since $x$ has its origin at the corotation
of the spiral, the curve of the spiral starts from (0,0). The warps cannot
propagate in the vertical forbidden band $[-\mu,+\mu]$, since their dispersion relation is
(in the infinitely thin disk limit) $\tomega_W^2=\mu^2+2\pi G\Sigma q$ (taking
into account the finite thickness of the disk and possible compressional effects
maintains the existence of such a forbidden band, see \eg Masset and Tagger 1995).
\new{For a given warp, the forbidden frequencies can be converted
into a forbidden band in $x$ as shown on figure for warp~2.}

We have depicted a narrow forbidden band since $\mu$ is expected to be small compared
to $\kappa$ if the rotation curve is nearly flat. In the ideal case where the rotation
curve is really flat, $\kappa=\sqrt{2}\Omega$ giving $\mu=0$.

According to the expression of system (I), we see that the coupling can be strong
where one or several group velocities vanish. Physically, this 
corresponds to the fact that the waves take a long time to propagate away 
from the region where coupling occurs, leaving them ample time to 
efficiently exchange energy and momentum.

\new{As already mentioned, we shall now use the group velocity
of a stellar spiral for the spiral wave, although we keep our purely 
gaseous coupling coefficient. By doing so, we avoid the tedious handling of a 
mixed kinetic-hydrodynamic
formalism without a great loss of precision.}
The group velocity of the spiral vanishes near the Lindblad resonances 
(here
at the Outer Lindblad Resonance, OLR) and the group velocity of the warp 
vanishes
at the edge of the forbidden band, where an incident warp must be reflected
\footnote{We can consider such a reflection as perfect, with
equal incident and reflected fluxes, and with vanishingly small transmission of another 
warp wave beyond the forbidden band, see Masset and Tagger 1995.}.

For these reasons we have four qualitatively distinct cases to consider:

{\bf Far from the OLR of the spiral:}
\begin{itemize}
\item{We consider two warps, none of which  close to its forbidden band;}
\item{Or a pair of warps, one of which lies at the edge of its 
forbidden band.}
\end{itemize} 
{\bf At the  OLR of the spiral:}
\begin{itemize}
\item{We consider two warps, none of which  close to its forbidden band;}
\item{Or a pair of warps, one of which lies at the edge of its 
forbidden band.}
\end{itemize} 

\subsubsection{Far from the OLR}

\begin{figure}
\psfig{file=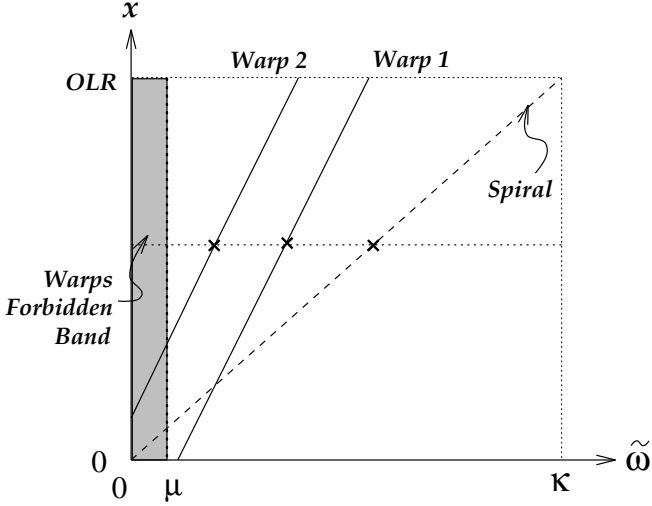,width=\columnwidth}
\caption{\label{fig:cas1} \scriptsize
This figure presents the first case studied in the text. We analyze the
efficiency of coupling at the radial position given by the horizontal 
dotted line, far from the OLR of the spiral (top of the diagram). Both warps are far 
from their
forbidden band.}
\end{figure}

Let us analyze the first case. The graph is presented in figure 
\ref{fig:cas1}. We choose a value of $x$ such that neither warp is close 
to its forbidden region. We define $H_t$ by $H_t^{1/2} = H_1^{1/2}+H_2^{1/2}$.  Summing 
the
two first lines of system (I), and after some transformations, we obtain:

\begin{equation}
\label{eqn:etape_1}
\partial_xH_t^{1/2} = \frac{\tomega_S\nu_Z^2}{2K_1K_2}\frac{1}{\sqrt{c_{g_1}c_{g_2}}}
\frac{\sigma}{\Sigma}H_t^{1/2}
\end{equation}

Our purpose is to determine whether the coupling will be sufficient in 
these conditions to extract
the warps from the noise level. As long as the warps remain at this low 
amplitude, their influence on the flux of the spiral is negligible.
We then assume that $\sigma/\Sigma$ is constant. This naturally leads us to
define an $e$-folding length for the warps. If this length happens to be 
short (in a sense which will be defined below), the warps can be
strongly amplified from the noise level.
Here, the  $e$-folding length is:

\[\lambda_e = \frac{2\sqrt{c_{g_1}c_{g_2}}K_1K_2}{\tomega_S\nu_Z^2}\frac{\Sigma}{\sigma}\]

Taking into account the expression of $K_i$ given in appendix \ref{apdx:energy}, and
noting that the group velocity of warp $i$ can be written as $c_{g_i}=
a\kappa/Q\tomega_i$,
we can rewrite this as:

\[\lambda_e = 4\frac{a}{Q\nu_Z^2}\sqrt{\biggl(1+\frac{\mu^2}{\tomega_1^2}\biggr)
\biggl(1+\frac{\mu^2}{\tomega_2^2}\biggr)}
\sqrt{\tomega_1\tomega_2}\frac{\Sigma}{\sigma}\]

where we have taken the typical value (see the diagram) $\tomega_S = \kappa/2$.

Now let us follow these two warps in their motion outward. After 
traveling 
over a distance $\lambda_e$, their quadratic total flux $H_t$ has been amplified
by a factor $e^2$. Will amplification at this rate be sufficient for the 
warps to extract a significant fraction of 
the flux of the spiral~? For this to occur, the spiral and the beat-wave 
of the warps 
should be able to maintain a well-defined relative phase as they travel 
radially, lest the coupling term oscillates and gives alternatively a 
positive and negative energy flux from the spiral to the warps. We meet 
here a condition which is the WKB equivalent of the selection rule on the 
radial wavenumbers: since the WKB wavenumbers of the waves do not a 
priori obey this selection rule, they will decorrelate over a few radial 
wavelengths, $k_x^{-1}\sim a/\Omega$, where $k_x$ is the radial 
wavenumber of any of the waves involved. Thus if $k_x\lambda_e$ is small 
the warps can be strongly amplified before they decorrelate from the 
spiral, whereas if $k_x\lambda_e$ is of the order of 1 or larger the 
coupling will not significantly affect them. Here, we have:

\[k_x\lambda_e = \frac{4}{Q}\underbrace{\frac{\Omega^2}
{\nu_z^2}}_{\sim .1}
\underbrace{\biggl(\frac{\tomega_1\tomega_2}{\Omega^2}\biggr)^2}_{\sim 1}
\underbrace{\frac{\Sigma}{\sigma}}_{\sim 10}
\sim \frac{4}{Q} \sim 1\]

Then the waves are in an adequate phase condition over one $e$-folding 
length at best, and we conclude that the warps cannot be extracted from 
the noise level in this first case.

\begin{figure}
\psfig{file=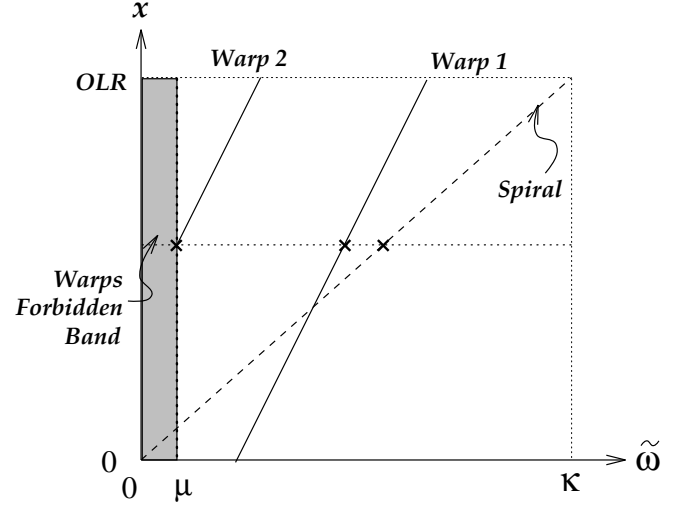,width=\columnwidth}
\caption{\label{fig:cas2} \scriptsize
This figure shows the second case studied in the text. We analyze the
efficiency of coupling at the radial position given by the horizontal 
dotted line, far from the OLR of the spiral (top of the diagram). Now one of the warps 
is located
just at the edge of its forbidden band.}
\end{figure}

The second case, illustrated in figure \ref{fig:cas2}, still corresponds 
to coupling 
far from the OLR of the spiral, but with one warp near the forbidden band. 
It is an easy matter
to check from the linear dispersion relation that the group velocity of a 
warp near its forbidden band is 
given by:

\begin{equation}
\label{eqn:modif2}
c_g(\Delta x) = \sqrt{\frac{A}{\mu}Q\kappa a\Delta x}
\end{equation}

where $\Delta x$ is the distance to the forbidden band. This formula has been obtained
in the WKB limit, but we do not expect very different results in the
general case. This leads us to replace in the expression of $\lambda_e$ a term
of the type: $H/\sqrt{c_{g_2}}$ by an integral of the type:

\[\int_0^H\frac{d(\Delta x)}{c_{g_2}^{1/2}(\Delta x)}\]

We have adopted the length scale $H$ because it corresponds to the range over which
equation~(\ref{eqn:modif2}) is valid, close to the forbidden band.

The ratio of these two quantities is:

\[\frac{4}{3}\biggl(\frac{\Omega\kappa}{A\mu}\biggr)^{1/4}\frac{1}{Q^{3/4}}\]

and is typically of the order of $1$ (the exponent $1/4$  does not allow 
it to change very much).
Furthermore, the coefficient $\sqrt{\tomega_1\tomega_2/\Omega^2}$ is now 
about
$\sqrt{\mu/\Omega}$, \ie about $1/2$.
This is not sufficient to make the value of $q\lambda_e$ much smaller than 
$1$.

Thus we can conclude that
far from its OLR the spiral cannot transfer its energy and angular momentum to the
warps.

\subsubsection{Near the OLR}
We turn now the last two cases, corresponding to coupling close to the
Outer Lindblad Resonance of the spiral. What does {\em close} mean here~?
We said that we must study separately the behavior of coupling near the OLR
because the group velocity of the spiral in a stellar disk tends to zero 
as its approaches 
the resonance.
In fact it does not really vanish (Mark, 1974), but we can in a first 
approximation 
consider that it does,
with a decay length scale of the order of the disk thickness. \new{Since 
this behavior is associated with the resonant absorption of the spiral wave at its 
OLR, by beating with the epicyclic motion of the stars, we expect that it 
should survive even in a realistic mixture of stars and gas.} 
From the dispersion relation for a stellar spiral given by Toomre (1969), 
and using
the asymptotic expression of Bessel functions, one can write the group 
velocity of the spiral near its OLR as:

\[c_{g_S} \simeq a\sqrt{Q}\biggl(\frac{4|A|}{\kappa}\biggr)^{3/2}\biggl(\frac{r_{OLR}
-r}{r}\biggr)^{3/2}\]

So that now the group velocity is a power $3/2$ of the distance to OLR.

We can slightly transform our expression of $\lambda_e$, the $e$-folding length of the
preceding section, as:

\[\lambda_e = \frac{4a}{Q\nu_z^2}
\sqrt{\tomega_1\tomega_2}\frac{\Sigma^{1/2}K_S}{H_S^{1/2}} c_{g_S}^{1/2}\]

This expression is more appropriate in this new situation, since (at least when
the fluxes of the warps remain small compared to the flux of the spiral) $H_S^{1/2}$ is constant
(while $\Sigma/\sigma$ is not) and now $c_{g_S}$ varies. As before, we have to see
how many  $e$-folding lengths are contained in a scale length of the 
order of $a/\Omega$, where we
can use the above expression for the spiral group velocity. Then we have
to replace $H/c_{g_S}^{1/2}$ in our previous estimate by $\int_0^Hdx/c_{g_S}^{1/2}(x)$,
leading us to multiply the number of $e$-folding lengths over a scale $H$ by:

\[\int_0^1 \frac{dx}{x^{3/4}}= 4 \]

Furthermore, we had taken in the previous section
the mean value $\tomega_S=\kappa/2$. Now since we are close to the OLR
we have to take
$\tomega_S = \kappa$. Finally, with our typical figures, instead of having 
only one $e$-folding 
length for 
$H_t^{1/2}$ over a scale 
length $H$, we find eight such $e$-folding lengths, \ie near the OLR
the flux of the warps can be multiplied by $e^{2\times8} = 10^7$. Of 
course this flux cannot become 
arbitrarily large, so that when it reaches the spiral flux we can consider 
that
most of the spiral flux has been transferred to the warps. Even though 
our factor of 
$10^7$ could be reduced by many effects, we consider that mode coupling 
near the OLR is efficient enough to allow the warps to grow from the noise 
level until they have absorbed a sizable fraction of the flux of energy 
and angular momentum carried by the spiral from the inner regions of the 
galactic disk.

{\new We now have to check in more details the effect of the 
localization of the coupling coefficient. Indeed from the system (I) 
we see that, if the amplitudes of the waves varied more rapidly than 
the coupling coefficient, their oscillations would reduce or even 
cancel over a given radial interval the efficiency of the coupling. 
This must be considered because the same effect which gives a 
vanishing group velocity for the spiral near its OLR also causes their 
radial wavenumber to diverge. 

However, from the expression above for the group velocity of the 
spiral, we see that:
\[\frac{\partial k_{r}}{\partial r}\sim \frac{\partial k_{r}}{\partial 
\omega}\sim (r_{OLR}-r)^{-3/2}\]
so that
\[k_{r}\sim (r_{OLR}-r)^{-1/2}\quad .\]
Now the phase of the spiral, in the WKB approximation, varies as
\[\exp{(\imath\int k_{r}dr)}\]
and the integral converges, so that in fact the phase tends to a 
finite limit (instead of varying rapidly as one might think from the 
divergence of $k_{x}$). Thus in the expressions on the RHS of system 
(I), the quantity which gives the dominant behaviour is indeed the 
(integrable) divergence of the coupling coefficient, in the vicinity of 
the OLR. This justifies that, in our estimates, we have integrated 
over the variations of this coefficient, keeping the other quantities 
approximately constant. It also justifies our statement, in section 
2.2, that the generation of the warps is impulsive-like. }

\begin{figure}
\psfig{file=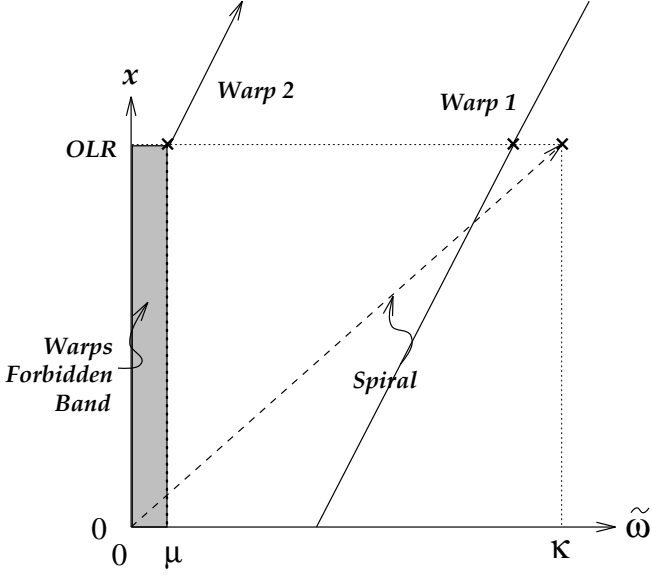,width=\columnwidth}
\caption{\label{fig:cas3} \scriptsize
This figure shows the final result of our study. The coupling is strongly
efficient at the OLR, and only there, and results in a conversion of the spiral
into two warp waves,  at $\tomega_2 = \mu$ and $\tomega_1 = \kappa-\mu$.}
\end{figure}

This does not tell us yet which couple of warps is preferentially 
amplified at OLR, according to their frequencies. 
In fact the coupling coefficient still depends on these frequencies, 
through the term in $\sqrt{\tomega_1\tomega_2}$. This factor, associated 
with the group velocities of the warps, further increases
the efficiency for the most ``external'' pair: 
$[\omega_1=\kappa-\mu,\omega_2=\mu]$ (by a factor $\sim 2$
compared to the most central pair of warps if we assume $\mu/\Omega \sim 
1/4$).
This difference is amplified geometrically at each $e$-folding 
length, and thus
finally we can consider that the pair $[\mu,\kappa-\mu]$ is preferentially  
emitted at
the OLR.
This result is illustrated in figure~\ref{fig:cas3}.

Now we have determined how and where the spiral is ``converted'' into 
warps. But we still have to find how the energy is distributed
between the warps, and in which direction they propagate. On figure 
\ref{fig:cas3} we have illustrated the warp 2 with an outward-pointing 
arrow, since we know
that it is necessarily emitted outward due to the presence of its 
forbidden band.
But warp 1 can consist of two waves, propagating inward as well as 
outward, and we must now consider the detailed balance
of the energy transferred to the three warps.
This can be done using the 
conservation of angular
momentum. The relation between the energy flux $H$ and angular 
momentum flux $J$ is, for the warps as well as the spiral (Collett and Lynden-Bell 1987):

\[J = \frac{mH}{\omega}\]

since angular momentum and energy are advected at the same velocity $c_g$.
Angular momentum conservation gives:

\[J_S = J_1^{(+)}+J_1^{(-)}+J_2\]

where $J_S$ is the momentum flux lost by the spiral, while $J_1,\ J_2$ 
are the fluxes gained by the warps, 
and the $(+)$ et $(-)$ upper-scripts denote respectively the warps 
1 emitted outward and inward. Writing $J_1=J_1^{(+)}+J_1^{(-)}$, and 
using the the fact that the waves 1 have the same frequency, we get 
$J_1=m_1H_1/\omega=H_1/\omega_1$, 
giving from energy and momentum conservation the system of equations:

\renewcommand{\arraystretch}{2.5}

\[ \left\{ \begin{array}{l}
\dss H_1 + H_2 = H_S
 \\
\dss H_1/\omega_1+H_2/\omega_2=2H_S/\omega_S
\end{array}
\right.
\]

\renewcommand{\arraystretch}{1}

Defining $\beta_1$ and $\beta_2$ as the ratios $H_1/H_S$ and $H_2/H_S$, 
we find
$\beta_i=\omega_i/\omega_S$.

Using the frequencies of the preferred pair of warps leads to:

\renewcommand{\arraystretch}{2.5}

\[ \left\{ \begin{array}{l}
\dss \beta_1 =  \frac{\Omega+\kappa-\mu}{2\Omega+\kappa}
\\
\dss \beta_2 = \frac{\Omega+\mu}{2\Omega+\kappa}
\end{array}
\right.
\]

\renewcommand{\arraystretch}{1}

A strictly flat rotation curve would give: $\beta_1=0.71$ and $\beta_2=0.29$,
while a nearly flat rotation curve with $\mu/\Omega = 0.25$, would give:
$\beta_1=0.63$ and $\beta_2=0.37$.

Now it is an easy matter to check that the fluxes carried away by warps $1^{(+)}$
and $1^{(-)}$ are equal. The easiest way to do it is to write for the pair
of warps~$1$ a system similar
to the one used above. This system is degenerate, but introducing
a slight offset between the frequencies and making it tend to zero we find
that they share equally the flux $\beta_1H_S$. Physically, this 
corresponds to the fact that the source of the waves is ``impulsional'' 
(very localized), so that they are emitted equally without a preferred 
direction in space.

From the numerical values given above, we can deduce that each of the three waves (two
emitted, one reflected) carries away one third of the flux extracted from 
the  spiral, as 
a very good approximation.

\subsection{Detailed physics at the OLR}
The situation exposed above is idealized. Two remarks modify
the simple computations, but should not qualitatively alter the
main physical conclusions.

\begin{itemize}
\item{First, the group velocity of the spiral does not
really vanish at the OLR. This has been investigated by Mark (1974),
who showed that the group velocity of the spiral at the OLR was about a few
kilometers per second (typically 10 times lower than the value it has
far from the OLR). This leads us to replace the factor $4$ in our
computation by about $\sim \sqrt{10}$. Nevertheless, we still have
a ``large'' number of $e$-folding lengths in the region ``just before'' the
OLR.}

\item{The spiral, as it approaches its OLR, is linearly damped through 
Landau effect. Mark (1974) gives an absorption length by Landau damping 
which has  
typically the same order of magnitude as the $e$-folding length we 
find for non-linear coupling. The two processes can thus compete, the 
spiral losing part of it energy and momentum flux to the stars (heating 
them and forming an outer ring), and to non-linearly generated warps, at 
comparable rates. It is thus very likely that the dominant process 
(non-linear coupling or Landau damping) will depend on the local 
characteristics of the galaxy under consideration.}

\end{itemize}

\section{Comparison with observations}

\begin{figure}
\psfig{file=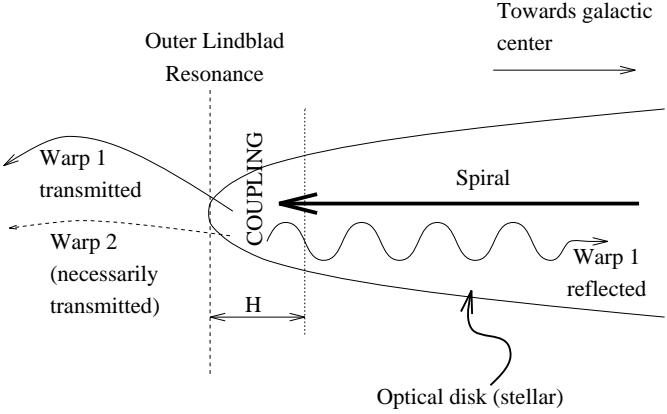,width=\columnwidth}
\caption{\scriptsize \label{fig:bilan}
This figure shows the spiral wave and the three
warp waves excited near its OLR by non-linear coupling. See the text for more details.}
\end{figure}

In this section we derive the expected characteristics
of the warps produced by the coupling process. Let us first summarize 
the main results obtained in the previous section. The reader
can also refer to figure \ref{fig:bilan} where the situation is
depicted.

\begin{itemize}
\item{The spiral wave travels outward until it reaches its OLR. Before it 
reaches it, the coupling with warps is too weak to affect the spiral.}
\item{When it reaches the OLR (more precisely within a distance of the 
order of $a/\Omega$ from the OLR) the spiral slows down and is
efficiently coupled with warps. This results in the ``conversion''
of the spiral into three warps which carry the  incident flux of the 
spiral 
away from the coupling region: two warps at frequency $\kappa-\mu$,
traveling respectively outward and inward, and
the last warp at frequency $\mu$, which is emitted outward.}
\item{The flux absorbed from the spiral is approximately equally shared by the three
warps, for reasons of conservation of angular momentum and energy.}
\item{In the following we will neglect the absorption of the spiral wave 
by Landau damping. Thus our results on the warps amplitudes will be
optimistic, but  still give results of the correct order of magnitude since the length
scale of Landau damping and non-linear coupling are similar.}
\end{itemize}

\subsection{Amplitude of the warp}
\new{We will now derive a rough prediction of the warp displacement 
resulting from our mechanism.}
For each of the three warps we can write:

\[c_{g_W}E_W \simeq \frac{1}{3}c_{g_S}E_S\]
where the factor $1/3$ comes from the result (see section 4.2.2) that the 
three warp waves share nearly equally the flux extracted from the spiral (\new{the rest of the flux of the spiral being absorbed
by Landau damping, as discussed in section~4.3}).

In order of magnitude, the flux of the spiral is given by:
\[H_S=c_{g_S}E_S\sim a \times a^2\Sigma_{disk}\biggl(\frac{\sigma}
{\Sigma_{disk}}\biggr)^2\]
Similarly, the fluxes of the warps emitted outward are:
\[H_W=c_{g_W}E_W\sim a \times \tomega_W^2\Sigma_{out}Z^2\]

In these expressions, $\Sigma_{disk}$ is the unperturbed surface density
of the stellar disk, where the spiral propagates, and $\Sigma_{out}$ is
the unperturbed surface density of the external HI disk, where the two 
emitted warps propagate.

\new{Now if we adopt a radial dependence of $\Sigma_{out}$ such as given by
Shostak and Van der Kruit (1984), that we will very roughly describe by~:

\[\Sigma_{out}(r) = \Sigma_0 \exp[-(r-r_0)/d]\]

where $r_0$ is the radius of the optical limit of the disk, $d$ the decaying 
length of surface density (a sizable fraction of $r_0$), we can deduce an
approximate radial dependence of the deviation $Z$. We find~:

\[\biggl({Z\over H}\biggr)^2 = \biggl(\frac{\sigma}{\Sigma_{disk}}\biggr)^2
\frac{\Sigma_{disk}}{3\Sigma_0}e^{(r-r_0)/d}\]

\[\Sigma_{disk}\sim 75 \;\;M_{\odot}.\rm{pc}^{-2}\]
and
\[\Sigma_{0}\sim 10^{-20}\mbox{part}.cm^{-2}\]

(see \eg Huchtmeier and Richter 1984 or Shostak and Van der Kruit 1984).}

and we take as a  value for the amplitude of the spiral 
\new{$\sigma/\Sigma_{disk} \sim 0.1$ (but it could be larger than this
widely adopted value, see Strom
\etal 1976).} This leads us to:

\new{
\[Z/H \sim {1 \over 2}e^{[(r-r_0)/2d]}\]

Of course $\Sigma_{out}$  decreases progressively  as the warps 
propagate outward, and $Z$ accordingly grows to preserve 
the flux.

The dependency found above is typical of what is observed in isolated spiral
galaxies.} It is noteworthy --- and we believe that
this has not yet been noticed --- that the flux of the spiral waves is 
of the same order of magnitude as the flux of the warps. We consider this 
coincidence as a strong point in favor of our mechanism ---~independently of 
its details ---, and a challenge to any alternate model.

\subsection{Expected characteristics of the corrugation}

The computations concerning the reflected warp are slightly different.
One should remember that it propagates in a region where both stars and 
gas are present, a fact which cannot be neglected when discussing its 
characteristics. In a previous paper (Masset and Tagger, 1995) we 
discussed the propagation of warps in a two-fluid disk. 
We found two types of waves: the first type corresponds to waves where 
both fluids move essentially together, while the second type corresponds 
to waves where the fluids move relative to each other; assuming that the 
stellar disk is much warmer and much more massive than the gaseous one, 
in this type of wave the stars stay essentially motionless while the gas 
moves in their potential well. We consider that this wave, which we call 
the corrugation mode, will be excited 
much more easily than the previous ones, since it moves much less mass and 
thus its amplitude  for a given energy 
must be much higher. Its dispersion relation is:

\[\tomega^2 = \mu^2+4\pi G\rho_*+2\pi G\Sigma_gq+\frac{\tomega}
{\tomega^2-\kappa^2}{a_g^2q^2}\]

where the subscripts $g$ and $*$ apply to gaseous and stellar values, and 
where $\rho_*$ is the stellar density at midplane.
Because of the moderate thickness of the stellar disk, the square 
frequency $\nu_z^2=4\pi G\rho_*$ is much larger than the other characteristic
frequencies of the problem, and in particular larger than the frequency 
$\tomega\simeq \kappa-\mu$ 
of the wave which can be non-linearly excited. A solution can 
nevertheless be found because inside the OLR of the warp the last term in 
the dispersion relation is negative, and can for large enough values of 
$q$ balance the dominant term $\nu_z^2$. 
Neglecting the self-gravity of the gas, and writing $\tomega^2 \ll 4\pi G\rho_*$ (a realistic
assumption since the ratio of these quantities is about one tenth), we find:

\[q = \frac{\sqrt{4\pi G\rho_*}}{a_g}\frac{\sqrt{\kappa^2-\tomega^2}}{\tomega}\]

Assuming a $sech^2(z/H)$ vertical stellar density profile (\ie the disk is
dominated by the local stellar gravity, which is a reasonable 
assumption), this can be rewritten as:

\begin{equation}
\label{eqn:lambda_corrug}
\frac{\lambda}{H_*} = 4.4 \frac{a_g}{a_*} \frac{1}{\sqrt{\kappa^2/\tomega^2-1}}
\end{equation}

Taking $a_g/a_*\sim .2$, and $H_*$ of the order of one kiloparsec, we find 
an order
of magnitude of one kiloparsec for the wavelength of the corrugation. This 
is consistent with the observed wavelength for the corrugation,  in 
the Milky Way (Quiroga
and Schlosser~1977, Spicker and Feitzinger~1986)
or in other galaxies (Florido \etal~1991a). Despite the observations by 
Spicker and Feitzinger of a
one kiloparsec corrugation in our galaxy, it seems that the typical 
wavelength of corrugation
is rather about 2~kiloparsecs, or even larger (Florido \etal~1991b). The 
discrepancy of $~2$ between
observations and our estimate could be due either to the magnetic field 
which affects the gas
motions and thus should imply the use of magnetosonic speeds instead of 
the acoustic
speed, or to a lack of resolution of some observations. 
From equation (\ref{eqn:lambda_corrug}), we find that the wavelength 
of corrugations 
increases as we approach the edge of the optical disk, in good 
agreement with observations.

Now let us give an order of magnitude of the expected amplitude of the 
corrugation. As already
explained, we can write that its flux is about one third of 
the
spiral flux. Now the flux of the corrugation is roughly given by (the index $c$ is for 
corrugation):

\[H_c = c_{g_c} E_c\]

where:

\[c_{g_c} = \biggl|\frac{\partial \tomega}{\partial q}\biggr| \simeq \frac{\tomega}{q}
\simeq \frac{\kappa}{q}=\frac{\kappa H_*}{qH_*} \sim a_*\]

since $qH_* \sim 1$, and:

\[E_c = \frac{1}{2}\nu_z^2\Sigma_gZ^2\]

Hence:

\[Z^2\sim\frac{\Sigma_*}{\Sigma_g}\frac{a_*^2}{\kappa^2} \biggl(\frac{\sigma}{\Sigma_*}\biggr)^2\]

which gives, taking into account $H_*/R \sim \kappa^2/\nu_z^2$:

\[Z=\sqrt{\frac{\Sigma_*}{\Sigma_g}\frac{H_*}{R}}H_*\biggl(\frac{\sigma}{\Sigma_*}\biggr)\]

Taking $H_* \sim 1\mbox{ kpc}$ and $\sigma/\Sigma_*\sim .1\mbox{ to }.2$, we obtain 
an expected amplitude for the corrugation of about
one hundred parsecs, which is typically the observed value. Note that, as 
discussed above,  
energy can also be
given to the other modes, involving both the gas and the stars, for the 
reflected warp. 
Their expected wavelength is larger 
than the disk radius, so that their line of nodes 
should be almost straight, corresponding 
to the observations of Briggs (1990), and so that the corrugation can easily be 
detected when
superimposed on such large scale bending waves.

\subsection{Straight line of nodes}
We have found thus far that our mechanism leads to results that fit well 
the observations for the amplitudes of the ``outer'' warps (the two waves 
emitted beyond the OLR of the spiral) and of the 
corrugation, and for the wavelength of the corrugation. We now turn to 
the wavelength of the outer warps. Actually the 
line of nodes of the observed warps is always nearly straight, leading to 
radial wavelengths larger than the galactic radius. This is known as the 
problem of the {\em straight line of nodes}, and is one of the most 
serious challenges to models of warps. In his work on the ``Rules of 
Behavior of Galactic Warps'', Briggs
(1990) has emphasized this remark: the line of nodes of the outside
warps appears nearly straight or with a slight spirality, always almost
leading. Later observations have shown that outside warps could also
be slightly trailing, but always nearly straight. This can be interpreted
from our mechanism, which predicts the emission of two warps outward
from the OLR. In fact, the frequency of the first warp is $\mu$, leading 
from the dispersion relation to a wavevector $k_x=0$, hence
a straight line of node. The second one is emitted at $\tomega\simeq\kappa-\mu$, \ie
near its OLR. From the dispersion relation of warps taking into account
the compressibility effects near the OLR (see \eg Masset and Tagger 1995),
we see that $k_x$ must tend to zero in order to balance the behavior of
the resonant denominator, thus also leading to a straight line of nodes.

Of course this is a qualitative and simple interpretation, since we 
use results obtained in the framework of the WKB assumption
(tightly wound waves) where $k_x$ is precisely not allowed to become small.
Hence this discussion about the straight
line of nodes is just the very beginning of a more complete work
which should take into account non-WKB effects and the cylindrical 
geometry. This work is in
progress, using the numerical code introduced in Masset and Tagger 1995.
Finally, we should mention that the combination of two warps outside
OLR can easily justify the large and arbitrary jump in the line of
nodes across the Holmberg radius observed by Briggs, since the relative phase
between warps 1 and 2 is arbitrary.
\section{Discussion}
\subsection{Summary of the main results}
We have written the non-linear coupling coefficients between a spiral and 
two warps, taking into account
the finite thickness of the galactic disk and the three dimensional motion 
in a warp. The efficiency of this mechanism is too 
weak
in the stellar disk, except at its outer edge where the spiral slows down 
near its OLR 
and
can be efficiently coupled to warps and extract them from the noise level, 
transferring to them a sizable fraction of 
its energy and angular momentum \new{(with the rest transferred to the 
stars by Landau damping)}. As a net result we can 
consider
that the spiral is transformed into two warps. One of them receives a third
of the energy of the spiral and can propagate only outward. The other receives the 
remaining two thirds
of the energy of the spiral, and splits into one wave traveling outward and one 
traveling inward, which we tentatively identify as the corrugation 
observed in many galaxies. 

Order of 
magnitude
estimates of the respective amplitudes of the outer warp (in the HI layer) 
and of the inner
one (the corrugation), as well as of their wavelengths, lead to values in 
good agreement with  
observations. In particular we find from the observed amplitudes that the 
energy and angular momentum 
fluxes of the spiral and the warps are comparable, a result which 
receives a 
natural explanation in our model.

The expected long wavelength of the outer warp is in agreement with the observed 
trend to forming 
a straight line of nodes. Furthermore, our mechanism justifies the phase discontinuity
of the warp at crossing the Holmberg radius (since the Holmberg radius is expected to
be close to the Outer Lindblad Resonance of the spiral, where our mechanism takes place).

\subsection{What about the halo ?}
In the preceding sections we have not mentioned the role of a massive 
halo. 
We will not discuss 
a  misalignment between the disk and the principal plane of the halo, 
since such
a misalignment was introduced in an {\it ad hoc} model by Sparke and 
Casertano (1984) in order to explain the continuous excitation of the 
warp in isolated galaxies. Our mechanism does not require  
such an hypothesis.

A typical halo is about ten times more massive than the disk. If this 
halo has a spherical symmetry, it does not affect the vertical restoring 
force and thus leaves our results unchanged. On the other hand a 
flattened halo would give rise to a strong restoring force
for a test particle (or a disk) displaced from the equatorial plane.
The vertical oscillation frequency $\mu$ would 
be about three times the epicyclic frequency. Hence the
forbidden band of figures \ref{fig:cas1}--\ref{fig:cas3} would 
invade more than the
whole disk. Our analysis would then have to be altered, but we believe 
that its main result would remain valid.

Indeed, in order to find two bending waves that would propagate in the 
vicinity of the OLR of the spiral, we might invoke the ``slow'' branch 
of the warp dispersion relation, which uses intensively the 
compressional behavior of the gas (let us emphasize that we have 
already made use of this slow branch to find the wavelength of 
corrugation, which is in the same manner a bending wave of the gas in 
the deep potential well of the stars).  This would not strongly change 
the expected efficiency of our mechanism, and would even rather 
increase it slightly, since we would be dealing with slower modes.  
Another solution would be to consider the pair of warps $[-\mu, 
\kappa+\mu]$, which are both simultaneously propagating (with the 
former being retrograde).  The efficiency of coupling with such a pair 
would be, in our framework (shearing sheet and steady state) exactly 
the same as the one we derived for the pair $[\mu, \kappa-\mu]$.  Now 
the details of the scenario would change a little.  We would have only 
one wave emitted outward and two inwards, but the order of magnitude 
of their amplitudes would not change.  Note that observations 
of corrugation in external galaxies (Florido \etal 1991a) seem to 
reveal two distinct wavelength for corrugation waves.

Another matter concerns the 
response of the
halo to the warp. A fully consistent treatment of 
this problem
should be numerical. However, it is possible to get analytical trends by 
considering
the halo as a thicker (and then warmer) disk than the stellar/gaseous 
disk. This can be done
using the two-fluid dispersion relation of warps (see Masset and Tagger 1995). 
The two-fluid
dispersion relation reveals three modes:

\begin{itemize}
\item{A mode which involves essentially the thinner and lighter 
component, leaving the other one (the halo) nearly motionless.  This 
mode is the one described in the preceding lines, and corresponds to 
the gaseous corrugation within the stellar disk as well as to the 
compressional mode of the whole disk (stars plus gas) in the rest 
potential of the halo.} \item{The classical two modes (slow and fast) 
of the one-fluid analysis, which involve similar motions of both 
fluids.  This means that the halo would participate in the different 
warps, as well as in the spiral, and formally the problem would be the 
same as the one described throughout this paper, with the difference that we would 
have to use the total surface density, involving the projected one of 
the halo, and the total thickness, which is the halo thickness, of the 
order of $r/2$.  Each of these quantities should then be multiplied by 
about ten.  Hence the test particle frequency $\nu_Z$ would remain the 
same, since it is proportional to $\sqrt{\Sigma/H}$.  We then see on 
equation~(\ref{eqn:etape_1}) that each factor would remain unchanged, 
except the group velocities whose expression would probably be 
slightly different, but could not differ much from the sound speed; 
hence the e-folding length would have the same behavior and order of 
magnitude.  Furthermore, a participating halo would give a more 
efficient and simple explanation to the problem of the straight line 
of nodes, since the wavelength of warps, for a given excitation 
frequency, increases with the surface density.}
\end{itemize}

\begin{figure}
\psfig{file=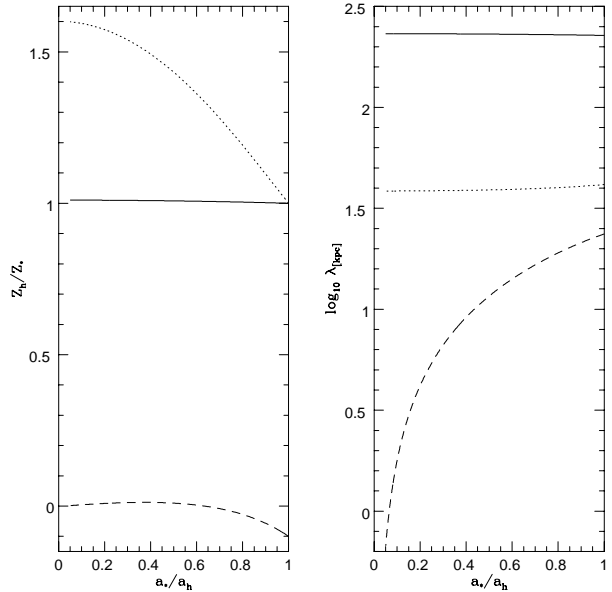,width=\columnwidth} \caption{\label{fig:halo} 
\scriptsize In this figure we present the three modes for bending 
waves in a two-fluids system (disk --denoted by a star-- and halo 
--denoted by a {\it h}--).  We have adopted the following parameters: 
$\Sigma_h/\Sigma_* = 10$, $\mu^2/\kappa^2 = 0.1$, $\tomega/\kappa = 
0.75$.  The results (the elongation $Z_{h}$ of the halo over the elongation  
$Z_{*}$ of 
the disk, and the wavelength) are presented as a function of the ratio 
$a_*/a_h$.  For a typical galaxy, this ratio should be about $0.2$.  
The solid line represents the classical infinitely thin one-fluid 
mode.  We see that its wavelength is far greater than the disk radius 
(hence it could be a beginning of the justification of the straight 
line of nodes problem) and that the disk and halo have nearly the same 
elongation.  The dotted line represents the one-fluid compressional 
mode.  Its wavelength is about four times the disk radius, and we see 
that the halo is more warped than the disk since the halo elongation 
is about $1.6$ that of the disk.  Finally the dashed curve represents 
the corrugation-like wave, \ie here the ``corrugation'' of the whole 
disk (stars and gas) in the rest potential of the halo, which appears 
to be in that case nearly motionless ( $Z_{h}\ll Z_{*}$).  }
\end{figure}

Figure \ref{fig:halo} shows that the participation of the halo to the 
global modes (\ie the
two waves corresponding to the one-fluid ones) is as strong as or 
stronger than that of the disk. Furthermore
the wavelengths of these modes agree with the observations of a straight 
line of nodes. However,
this figure has been obtained in the WKB assumption and moderately thick 
disk formalism
(Masset and Tagger 1995), which are both
far too rough in this case, so that a numerical solution 
taking into account
the halo response would be necessary.

In the same manner, if we want to estimate the coupling efficiency taking 
into account the halo,
a problem is a determination of $\sigma/\Sigma$ for the spiral
involving a participating halo. If we assume a behavior of matter independent of $z$, 
then $\sigma/\Sigma$
should still be about~$0.2$, but once again only a numerical solution would allow us 
to give the real vertical profile
of perturbed quantities in the spiral and in the warps. Furthermore, 
the problem is much more
complex if we want to treat it fully correctly, since for distances to the 
galactic center
less than the Holmberg radius we have to cope with {\it three} fluids: the 
halo, the stars and
the gas, so the method which consists in treating the disk as a unique 
fluid will give only
approximate results.

Nevertheless, despite the large number of allowed modes of propagation of 
bending waves
in this realistic case, we see that global modes (\ie involving a motion 
of the halo) will
lead to the same efficiency as found in this paper. Numerical 
simulations would be needed to give an accurate evaluation of the 
e-folding length and to determine
the fraction of incident energy that each wave carries away.

Finally, let us mention that the halo should be a collisionless fluid 
(\ie star-like and not gas-like) 
if we still want the spiral
to slow down at the OLR. Observations suggesting that the halo or outer 
bulge is composed
of very low mass stars would confirm this hypothesis (see e.g. Lequeux 
\etal 1995).

\subsection{Suggested observations for checking the spiral-warps coupling mechanism}
In this paper we have shown that the spiral wave is converted into bending waves observed
as an outer HI warp and a corrugation wave. We have already emphasized the observational
fact that the fluxes of these three waves have the same order of magnitude for a typical
galaxy. It would be interesting to test  this coincidence with a better accuracy, and to search
for a correlation between the fluxes of the spiral, the warp and the corrugation. The fluxes
can be straightforwardly obtained from such observables as the disk surface density,
the rotation curve (which gives values for $\kappa$ and $\mu$), the disk thickness,
the perturbed surface density
of the spiral (see e.g. Strom \etal~1976), and the deviation of the disk to the equatorial
plane --- although some of these quantities are better measured in face-on 
galaxies, and others in edge-on ones, making a complete set of 
observations in an individual galaxy a challenge. A correlation between the fluxes for a significant number of galaxies would be a
strong argument for our mechanism.

\appendix

\section{Formal derivation of the coupling coefficient}

\label{apdx:formel}

\subsection{Derivation without the shear terms}

We use the hydrodynamical equations (continuity and Euler) and the Poisson
equation.We write the first two equations in tensor formalism, in
order to avoid too lengthy expressions. The continuity equation is:

\begin{equation}
\label{eqn:conti0}
\partial_t\rho+\partial^i(\rho V_i) = 0
\end{equation}

where the vector $V_i$ ($i=1,2,3$) is a tensor which represents
$(U,V,W)$, and where we implicitly sum over repeated indices (Einstein's
convention).

The Euler equation is:
\begin{equation}
\label{eqn:euler0}
\frac{\partial V^i}{\partial t}+V^j\partial_jV^i+2\epsilon^{ijk}\Omega_jV_k=
-a^2\frac{\partial i [\rho_0 (1+s)]}{\rho_0 (1+s)}-\partial^i\phi
\end{equation}

In this expression the last term of the L.H.S. represents the Coriolis
acceleration, and $\epsilon^{ijk}$ the Levi-Civita symbol. We also introduce
a convenient notation, $s=\rho/\rho_0$. The reader can note that, in
the convective derivative $(V.\nabla)V$, we have not written the shear
terms. We neglect them in this first approach, and we will check their
influence at the end of this section.

In the Euler equation the pressure term can be expanded to second
order as:

\[-a^2\biggl[(1-s)\partial^is+\frac{\partial^i\rho_0}{\rho_0}\biggr]\]

Finally, the Poisson equation is:
\begin{equation}
\label{eqn:poisson0}
\Delta \phi = 4\pi G\rho
\end{equation}

We want to know the energy injection rate into warp 1 from the spiral
and the warp 2. In the linear regime, we would have: $\partial_tE_1=0$. Here,
we will have a non-vanishing expression due to the presence of the spiral
and warp 2.

Since we treat the warp 1, we expand the hydrodynamic equations to second 
order in the perturbed quantities and project them onto
$e^{-i\omega_1 t}$.

For the continuity equation (\ref{eqn:conti0}) this gives:

\begin{equation}
\label{eqn:conti1}
\frac{\partial \rho_1}{\partial t}+\partial_i(\rho_0 V_1^i)
+\partial_i(\rho_S V_2^{i*})+\partial_i(\rho_2^* V_S^i) = 0
\end{equation}

and for the Euler equation (\ref{eqn:euler0}):

\begin{equation}
\label{eqn:euler1}
\frac{\partial V_1^i}{\partial t}+V^j_S\partial_jV_2^{i*}
+V_2^{j*}\partial_jV_S^i+2\epsilon^{ijk}\Omega_jV_{k_1}=
\end{equation}
\[-a^2\partial^is_1-\partial^i\phi_1+a^2s_S\partial^is_2^*
+a^2s_2^*\partial^is_S\]

Our purpose is to write the temporal derivative
of an expression which can be interpreted as the energy density of the
warp (energy density per unit surface).
For this we multiply equation (\ref{eqn:euler1}) by $a^2\rho_1^*/\rho_0$
and equation (\ref{eqn:conti1}) by $\rho_0$, we integrate over $z$, add them
and their complex conjugates. This gives:

\begin{equation}
\label{eqn:coupstep1}
\partial_t\int[\rho_0(V_1^iV_{1_i}^*+a^2s_1s_1^*)]+
\biggl[\int\rho_0V_{i_1}\partial^i\phi_1^*+c.c.\biggr]
\end{equation}

\[=-\int a^2\rho_0V_{i_1}^*\partial^is_1-\int a^2s_1\partial_i(\rho_0V_1^{i*})
-\int\rho_0V_{1_i}^*V^j_S\partial_jV^{i*}_2\]
\[-\int\rho_0V^*_{1_i}V^{j*}_2\partial_jV^i_S
+\int\rho_0a^2V^*_{1_i}s_S\partial^is^*_2+\int\rho_0a^2V^*_{1_i}s_2^*\partial^is_S\]
\[-\int a^2s_1^*\partial_i(\rho_SV^{i*}_2)-\int a^2s^*_1\partial_i(\rho_2^*V^i_S)+c.c.\]

In this tensorial notation we will freely use the rule of integration by part (\ie transpose 
derivative operators with a change of sign) in the linear terms of 
this equation: indeed if the index $i$ is $z$, this is really an 
integration by parts; on the other hand if $i=x$ or $y$ the derivative is just 
a multiplication by $ik_{i}$ (in our WKB analysis), so that \eg:
\[s_{1}\partial_{i}(\rho_{0}V_{1}^{i*})=-\rho_{0}V_{1}^{i*}\partial_{i}s_{1}\quad .\]
Furthermore, we must remember that the integration is only over $z$, 
so that we do not need to worry about boundary terms the integration 
by parts would introduce in an 
integration over $x$, and which would represent energy and momentum 
flux at the radial boundaries of the integration range.
  
However, we cannot blindly apply this simple formal integration by 
parts on the non-linear terms.  A careful look at the details of the 
integration by parts for each value of the index $i$ is necessary.  
For the dimensions ($i\equiv y$ and $i\equiv z$) the formal 
integration by parts would give correct results, even on the 
non-linear terms.  But for $i\equiv x$, an integration by part would 
imply a selection rule of the type $k_{x_S} = k_{x_1}+k_{x_2}$, and we 
have intentionally avoided such an hypothesis.  Since we want to stay, 
for simplicity, in the framework of tensorial formalism, we have to 
perform transformations that are correct whatever the index $i$ and we 
do not integrate by parts the non-linear terms in this first part of 
the derivation.

Equation (A6) already has the form we want : the first term in its L.H.S.
is the temporal derivative of the kinetic and internal energy of warp 1.
We can note that the term linked to Coriolis acceleration has vanished,
since the Coriolis force is always normal to the velocity and thus does not work.

The second term in the L.H.S. must be rewritten to be interpreted as
the temporal derivative of the potential energy. We integrate
it by parts and then use equation (\ref{eqn:conti1}) to transform the
result. We can also note that the first two terms of the R.H.S vanish
together (integrating one of them by parts). We obtain:

\begin{equation}
\label{eqn:coupstep2}
\partial_t\int[\rho_0(V_1^iV_{1_i}^*+a^2s_1s_1^*)]+\biggl[
\int\phi_1\partial_t\rho_1^*+c.c.\biggr]
\end{equation}

\[=-\int\rho_0V_{1_i}^*V^j_S\partial_jV^{i*}_2-\int\rho_0V^*_{1_i}V^{j*}_2\partial_jV^i_S
+\int\rho_0a^2V^*_{1_i}s_S\partial^is^*_2\]
\[+\int\rho_0a^2V^*_{1_i}s_2^*\partial^is_S
-\int a^2s_1^*\partial_i(\rho_SV^{i*}_2)-\int a^2s^*_1\partial_i(\rho_2^*V^i_S)\]
\[+\int\phi_1^*\partial_i(\rho_SV_2^{i*})+\int\phi_1^*\partial_i(\rho_2^*V_S^i)+c.c.\]

The second term in the L.H.S still does not appear as the time derivative of
a potential energy. We have to use the Poisson equation, which we have not used yet.

If we write $S(x^i-x^i_0)$ for the potential created  at point $x^i$
by a unit mass located at point $x^i_0$, then:

\[\phi = S*\rho\]

where $*$ represents the convolution operator.

Thus, $\phi_1=S*\rho_1$, $\phi_S=S*\rho_S$, etc\ldots.

We write $\cal L(\rho,\rho')$ the form:

\[{\cal L}(\rho,\rho') = \int_E\rho^*(\vec r)[S*\rho'](\vec r)d^3\vec r\]

where the index $E$ indicates that the integral extends over
the whole space. Using the fact that $S$ is a real function with spherical symmetry,
it is an easy matter to check that (by writing explicitly the convolution
and exchanging the integrals):

\[{\cal L}(\rho,\rho') = {\cal L}(\rho',\rho)^*\]

Hence $\cal L$ is a variational form, and one can write the following
equalities:

\[\int\phi_1^*\partial_t\rho_1={\cal L}(\partial_t\rho_1,\rho_1)
={\cal L}^*(\rho_1,\partial_t\rho_1)\]

(whether the integral extends only over $z$ or over the whole space,
since the integrated quantity does not depend on $x$ or $y$).

Now:

\[S*\partial_t\rho_W = \partial_t(S*\rho_W)\]

since $S$ is time independent. This is linked to the use of the Poisson
equation, \ie to the fact that the potential propagates instantaneously.
Then one can write:

\[\int\phi_1\partial_t\rho_1^* = \int\rho_1^*\partial_t\phi_1\]

Thus we see that we can write this term as a potential energy associated
to the warp, and this gives:

\begin{equation}
\partial_t\int\rho_0\biggl[(V_1^iV_{1_i}^*+a^2s_1s_1^*)+\biggl(\frac{1}{2}
s_1^*\phi_1+c.c.\biggr)\biggr]
\end{equation}

\[=-\int\rho_0V_{1_i}^*V^j_S\partial_jV^{i*}_2-\int\rho_0V^*_{1_i}V^{j*}_2\partial_jV^i_S
+\int\rho_0a^2V^*_{1_i}s_S\partial^is^*_2\]
\[+\int\rho_0a^2V^*_{1_i}s_2^*\partial^is_S
-\int a^2s_1^*\partial_i(\rho_SV^{i*}_2)-\int a^2s^*_1\partial_i(\rho_2^*V^i_S)\]
\[+\int\phi_1^*\partial_i(\rho_SV_2^{i*})+\int\phi_1^*\partial_i(\rho_2^*V_S^i)+c.c.\]

Hence we have expressed the temporal derivative of the energy density 
(kinetic plus internal plus gravitational) $E_1$ of warp 1 
 as a sum of integrals which involve
the amplitude of each warp and of the spiral. In particular, one may note that
in the linear case the energy of the warp is conserved. One important remark
concerning the linear case is that, if we neglect the coupling term, we should not
have $\partial_tE_1 = 0$ but $\partial_tE_1+\partial_x(c_gE_1)=0$ since energy is
transported at the group velocity $c_g$. Here the convective term has 
disappeared in the integration by parts of one of the terms in the RHS of 
equation (A6): for simplicity we have neglected the integrated terms, assuming periodicity 
in $x$. From a more complete derivation we would recover a linear term 
$\partial_x(c_gE_1)$, without affecting the non-linear ones. We will thus 
re-write (A8) as:

\[\partial_tE_1 + \partial_x(c_gE_1) \equiv \frac{d}{dt}E_1 = \mbox{ Coupling Term}\]

\subsection{Effect of the shear term}
As emphasized above, we did not take into account the shear term when writing
down the Euler equation. Writing this term can {\it a priori} modify the
warp energy and the coupling term.

Formally, in tensorial notation, the shear can be written as:
\[V^i_0 = 2A^{ij}x_k\]

where the matrix {\boldmath $A$} is, in the frame $xyz$:

\[\left( \begin{array}{ccc}
 0      &       0       &       0 \\
 A      &       0       &       0 \\
 0      &       0       &       0 
\end{array}
\right)
\]

With these notations, the convective term of the Euler equation reads:

\[(2A^{jk}x_k+V^j)\partial_j(2A^{il}x_l+V^i)\]

which can be rewritten as:

\[4A^{jk}A^i_jx_k+2A^i_jV^j+2A^{jk}x_k\partial_jV^i+V^j\partial_jV^i\]

The first of these four terms, proportional to  $\mbox{\boldmath $A$}^2$, is null
(this is generally the case when the velocity is constant along a current line).

The last term is the one we had without shear. The influence of this term
on the coupling coefficient has already been derived in the previous section.

The intermediate terms are linear, and thus have no incidence on
the coupling term.

Hence, the temporal derivative of the energy density of the warp is modified
by the addition of the two corresponding terms, obtained by multiplying
by $V^*$ and adding the complex conjugates. This gives:

\[\partial_tE \longrightarrow
\partial_tE+4\Re[\int\rho_0V_i^*A^i_jV^j]+4\Re[\int\rho_0V_i^*A^{jk}x_k\partial_jV^i]\]

where \(\Re\) stands for the real part. For the waves under consideration both of these 
terms are null:
\begin{itemize}
\item{The first one because, in the WKB approximation, the perturbed horizontal
velocities $U$ and $V$ are in quadrature (epicyclic motion). Thus this term, given the
expression of {\boldmath $A$}, is proportional to $UV^*$, and therefore purely
imaginary.}
\item{The second one because it is related to the azimuthal derivative of terms
related to the warp  energy density, and thus vanishes.}
\end{itemize}

Hence the addition of shear does not modify the coupling equation.

\section{Analytical computation of the coupling term}

\label{apdx:vecppe}

\subsection{Approximation of the eigenfunctions}

In the 
framework of our expansion to second order in the perturbed quantities 
(assuming weak non-linearities), 
the coupling integrals derived in the preceding appendix involve the {\it linear} 
eigenfunctions associated with the spiral and the warp. These 
eigenfunctions consist in the amplitude and phase relation between the 
perturbed quantities (velocities, density, potential...), and their 
spatial variations. An exact, analytical or numerical, knowledge of these 
eigenfunctions is beyond the scope of this work, and in fact would add 
little to it since we are not interested here in 
deriving detailed numbers appropriate to a given galaxy model but rather 
in the physics of non-linear mode coupling. We will thus use approximate 
expressions, which allow an easier access to this physics. Furthermore, 
our results are given in terms of variational forms. These are well known 
to preserve the important invariants in the problem (the energy and 
action densities), and to be good estimates even with poor approximations 
to the eigenvectors. 

Thus we remain in the WKB formalism, and make the following
assumptions:
\begin{itemize}
\item{We assume that the motion in the spiral is purely horizontal and
independent of $z$.}
\item{We assume that the vertical velocity in warps is independent of $z$, and that
the perturbed density is: \[\rho_W = -Z_W\partial_z\rho_0\]}
\end{itemize}

From these assumptions we can deduce the perturbed quantities relative to a warp.
We focus hereafter on the warp 2, but the results would evidently apply 
also to warp 1.

One gets $U_2$ et $V_2$, the horizontal components of the perturbed velocity, from the relation:
\[k_{x_2}U_2+k_{y_2}V_2=
\frac{q_2^2\tomega_2}{\tomega_2^2-\kappa^2}(\phi_2+a^2s_2)\]

Using the WKB hypothesis ($k_{y_2} \ll k_{x_2} \simeq q_2$), one obtains:

\[U_2 = \frac{q_2\tomega_2}{\tomega_2^2-\kappa_2^2}(\phi_2+a^2s_2)\]
\[V_2 = \frac{2i\Omega}{\kappa}\frac{q_2\tomega_2}
{\tomega_2^2-\kappa_2^2}(\phi_2+a^2s_2)\]

where $\phi_2$ is known from $\rho_2$. 

It is noteworthy that these quantities are only 
a rough approximation to the  eigenfunction of a warp; in particular they do not fulfill
the continuity equation. On the other hand our reason for taking them 
into account is that 
near the Lindblad resonances, where horizontal motions are large, we 
suspect that they might strongly contribute to the coupling terms. We 
will find later that this is not the case, so that forgetting these terms 
altogether would not change the result to leading order.

The perturbed potential of the warp is given by the expansion to lowest order
in $qH$ of the expression given by the Green functions (see Masset and Tagger 1995):

One obtains :

\[\phi_2(z) = -4\pi GZ_2\int_0^z\rho_0(z')dz'\]

and then the other perturbed quantities associated with the warp can be expressed, as functions
of $Z_2$ exclusively:

\[U_2 =  -\frac{q_2\tomega_2}{\tomega_2^2-\kappa_2^2}
\biggl(4\pi G\int_0^z\rho_0(z')dz'+a^2\frac{\partial_z\rho_0}{\rho_0}\biggr)Z_2\]

\[V_2 =  -\frac{2i\Omega}{\kappa}\frac{q_2\tomega_2}{\tomega_2^2-\kappa_2^2}
\biggl(4\pi G\int_0^z\rho_0(z')dz'+a^2\frac{\partial_z\rho_0}{\rho_0}\biggr)Z_2\]

\[W_2 = -i\tomega_2Z_2\]

\[s_2 = -Z_2\frac{\partial_z\rho_0}{\rho_0}\]

In the same manner, with the hypotheses mentioned above, we obtain for 
the spiral:

\label{fonctions_propres}

\[U_S = \frac{\tomega_S}{q_S}\frac{\sigma}{\Sigma}\]

\[V_S = \frac{2i\Omega}{\kappa} \frac{\tomega_S}{q_S}\frac{\sigma}{\Sigma}\]

\[W_S = 0\]

\[\phi_S = -\frac{2\pi G\sigma}{q_S}\]

\[s_S = \frac{\sigma}{\Sigma}\]

Rewriting the coupling term obtained in the previous section, and expanding
the sums over repeated indices, we get:

\begin{equation}
\label{eqn:couplform2}
\frac{d}{dt}E_1=\int i(q_S-q_2)\rho_0\phi_1U_2s_S^*+\rho_0s_S^*W_2\partial_z\phi_1
\end{equation}

\[+i(q_S-q_2)\rho_0U_S^*s_2\phi_1+i(q_2-q_S)\rho_0a^2s_S^*(U_1s_2-U_2s_1)\]
\[+\rho_0a^2s_S^*(W_1\partial_zs_2+W_2\partial_zs_1)
+i(q_S-q_2)\rho_0a^2s_1s_2U_S^*\]
\[+i\rho_0U_1U_2U_s^*(q_s-q_2)
-\rho_0iq_2V_1V_2U_S^*-\rho_0iq_2W_1W_2U_S^*\]
\[+\rho_0iq_SU_2V_1V_S^*+c.c.\]

The final step consists in replacing the quantities involved in these integrals
by the approximate perturbed quantities derived above.

\subsection{Computation of the coupling integrals}

In this appendix we compute the various terms in the RHS of the coupling 
equation (\ref{eqn:couplform2}).

We will not write down all the computations, which are lengthy and tedious.
Let us just notice that the spiral eigenfunctions do not depend on $z$,
so that each integral over $z$ involves in fact only two eigenfunctions 
of warps 1 and 2.

We will make intensive use of integrals of the type:
\[I_1 = \int_{-\infty}^{+\infty}a^2\frac{(\partial_z\rho_0)^2}{\rho_0}dz\]
and
\[I_2 = \int_{-\infty}^{+\infty}\biggl[\partial_z\rho_0\int_0^z\rho_0(z')dz'\biggr]dz\]

All the other integrals are either straightforward or can be deduced from $I_1$
or $I_2$.

For the evaluation of $I_1$, we use the hydrostatic equilibrium in the 
unperturbed state:

\[a^2\partial_z\rho_0 = -\rho_0\partial_z\phi_0\]

Hence
\[I_1 = -\int_{-\infty}^{+\infty}\partial_z\phi_0\partial_z\rho_0dz\]
Integrating by parts, one obtains:

\[I_1 = \int_{-\infty}^{+\infty}\rho_0\partial_{z_2}^2\phi_0\]
Now
\[\partial_{z_2}^2\phi_0 = \Delta\phi_0 - \Delta_r\phi_0\]
where $\Delta_r\phi_0$, the radial laplacian of $\phi_0$ is  equal to $-\mu^2$
(Hunter and Toomre, 1969).

Using the Poisson equation, $\Delta\phi_0 = 4\pi G\rho_0$, one obtains:

\[I_1 = 4\pi G \Sigma_2 + \mu^2\Sigma\]
where $\Sigma_2$ is defined as:

\[\Sigma_2 = \int_{-\infty}^{+\infty}\rho_0(z)^2dz\]

Using an integration by parts, one can easily derive:
\[I_2 = -\Sigma_2\]

The derivation needs a frequent use of the following integral:

\[I_3 =  \int_{-\infty}^{+\infty}\rho_0\biggl[4\pi G\int_0^z\rho_0(z')dz'+
a^2\frac{\partial_z\rho_0}{\rho_0}\biggr]^2dz\]

Expanding the square in the integral:

\[I_3 =  \int_{-\infty}^{+\infty}16\pi^2G^2\rho_0\biggl(\int_0^z\rho_0(z')dz'\biggr)^2
+a^2I_1+8\pi Ga^2I_2\]

One can deduce:

\[I_3 = \frac{4\pi^2 G^2\Sigma^3}{3} - 4\pi Ga^2\Sigma_2 +\mu^2a^2\Sigma\]

In order to express the coupling term , we group the terms as follow:

\begin{equation}
\frac{d}{dt}E_1=\int \underbrace{i(q_S-q_2)\rho_0\phi_1U_2s_S^*}_{T_1}+
\underbrace{i\rho_0s_S^*W_2\partial_z\phi_1}_{T_2}
\end{equation}

\[+\underbrace{i(q_S-q_2)\rho_0U_S^*s_2\phi_1}_{T_3}
\underbrace{+i(q_S-q_2)\rho_0a^2s_S^*(U_1s_2-U_2s_1)}_{T_4}\]
\[+\underbrace{i\rho_0a^2s_S^*(W_1\partial_zs_2+W_2\partial_zs_1)}_{T_5}
+\underbrace{i(q_S-q_2)\rho_0a^2s_1s_2U_S^*}_{T_6}
\]
\[+\underbrace{i\rho_0U_1U_2U_s^*(q_s-q_2)}_{T_7}
\underbrace{-\rho_0iq_2V_1V_2U_S^*}_{T_8}
\underbrace{-\rho_0iq_2W_1W_2U_S^*}_{T_9}\]
\[+\underbrace{\rho_0iq_SU_2V_1V_S^*}_{T_{10}}+c.c.\]

and we find for these terms the expressions:

\renewcommand{\arraystretch}{2.5} 

\[ \left\{ \begin{array}{lcl}
T_1 & = & \dss\frac{(q_S-q_2)q_2\tomega_2}{\tomega_2^2-\kappa^2}\biggl[\frac{
4\pi^2G^2\Sigma^2}{3}-4\pi Ga^2\frac{\Sigma_2}{\Sigma}\biggr]Z_1Z_2\sigma \\
T_2 & = & \dss 4\pi G\tomega_2\frac{\Sigma_2}{\Sigma}Z_1Z_2\sigma \\
T_3 & = & \dss 4\pi G\tomega_s\frac{q_S-q_2}{q_S}\frac{\Sigma_2}{\Sigma}Z_1Z_2\sigma \\
T_4 & = & \dss a^2\mu^2(q_S-q_2)\biggl(\frac{q_1\tomega_1}{\tomega_1^2-\kappa^2}
-\frac{q_2\tomega_2}{\tomega_2^2-\kappa^2}\biggr)Z_1Z_2\sigma
\\
T_5 & = & \dss -\biggl(4\pi
 G\frac{\Sigma_2}{\Sigma}+\mu^2\biggr)(\tomega_1+\tomega_2)Z_1Z_2\sigma
\\
T_6 & = & \dss \frac{(q_S-q_2)\tomega_S}{q_S}\biggl(4\pi
 G\frac{\Sigma_2}{\Sigma}+\mu^2\Sigma\biggr)Z_1Z_2\sigma
\\
T_7 & = & \dss \frac{q_S-q_2}{q_S}\tomega_S
\frac{q_1q_2\tomega_1\tomega_2}{(\tomega_1^2-\kappa^2)(\tomega_2^2-\kappa^2)}
\\
& & \dss \times
\biggl(\frac{4\pi^2G^2\Sigma^2}{3}-4\pi Ga^2\frac{\Sigma_2}{\Sigma}+\mu^2a^2\biggr)
Z_1Z_2\sigma
\\
T_8 & = & \dss \frac{q_2}{q_S}\tomega_S
\frac{4\Omega^2}{\kappa^2}
\frac{q_1q_2\tomega_1\tomega_2}{(\tomega_1^2-\kappa^2)(\tomega_2^2-\kappa^2)}
\\
& & \dss \times
\biggl(\frac{4\pi^2G^2\Sigma^2}{3}-4\pi Ga^2\frac{\Sigma_2}{\Sigma}+\mu^2a^2\biggr)
Z_1Z_2\sigma
\\
T_9 & = & \dss \frac{q_2}{q_S}\tomega_S\tomega_1\tomega_2Z_1Z_2\sigma
\\
T_{10} & = & \dss \tomega_S
\frac{4\Omega^2}{\kappa^2}
\frac{q_1q_2\tomega_1\tomega_2}{(\tomega_1^2-\kappa^2)(\tomega_2^2-\kappa^2)}
\\
& & \dss \times
\biggl(\frac{4\pi^2G^2\Sigma^2}{3}-4\pi Ga^2\frac{\Sigma_2}{\Sigma}+\mu^2a^2\biggr)
Z_1Z_2\sigma
\end{array}
\right.
\]

\renewcommand{\arraystretch}{1} 

We note that the coefficients in eq.  (B2) are all imaginary, and 
that all the expressions above give a real factor times 
$Z_{1}Z_{2}\sigma$.  Thus if the relative phases of the three waves 
were such that this product is real the coupling term (once added  
to its complex conjugate) would exactly vanish, while it would be 
maximized by waves in quadrature. This variation with the phases is 
associated with the complex non-linear behaviors mentioned earlier, 
and we will not discuss it here (technically the phases can evolve 
non-linearly on the ``slow'' time scale of the mode growth and 
non-linear evolution, as compared with the ``fast'' time scale of the 
linear frequency). Hereafter we will for simplicity 
assume that the non-linear process has picked up waves from the background 
noise, or made them evolve, such that their phases maximize the 
coupling, \ie $Z_{1}Z_{2}\sigma$ imaginary. Thus we will from now only 
consider real quantities in the coupling equation.

Factorizing $\sum_{i=1}^{10}T_i$, we obtain the following coupling equation:

\begin{equation}
\label{eqn:coupldetail}
\frac{d}{dt}E_1= \Biggl[
\beta\Theta_1\Theta_2\frac{q_1q_2}{\kappa^2}\tomega_S\bigg[1+\frac{4\Omega^2}{\kappa^2}
+\biggl(1-\frac{4\Omega^2}{\kappa^2}\biggr)\frac{q_2}{q_S}\biggr]
\end{equation}

\[+(\tomega_1\tomega_2+\eta^2-2\mu^2)\frac{q_2}{q_S}\tomega_S +
(\mu^2-\eta^2)
\biggl(\tomega_2+\frac{q_S-q_2}{q_S}\tomega_S\biggr)\]
\[+\frac{q_S-q_2}{\kappa}[q_2\Theta_2(\beta+a^2\mu^2)-q_1\Theta_1a^2\mu^2]
\Biggr]Z_1Z_2\sigma
\]

where we have introduced:

\renewcommand{\arraystretch}{2.5}

\[ \left\{ \begin{array}{l}
        \dss \eta^2=\mu^2-4\pi G \Sigma_2/\Sigma \\
        \dss \beta =  4\pi^2G^2\Sigma^2/3 + a^2\eta^2 \\
        \dss \Sigma_2 = \int_{-\infty}^{+\infty}\rho_0(z)^2dz \\
        \dss \Theta_1 = \tomega_1\kappa/(\tomega_1^2-\kappa^2) \\
        \dss \Theta_2 = \tomega_2\kappa/(\tomega_2^2-\kappa^2) 
        \end{array}
   \right.
\]

\renewcommand{\arraystretch}{1}

We clearly see that we are concerned with inhomogeneous coupling since the
coefficient depends on $x$, which implicitly comes from the $\tomega$.
In particular, the dimensionless coefficients $\Theta_1$ and
$\Theta_2$ are of the order of unity,
except near the Lindblad resonances of the corresponding warps where they
can take very large values.

\section{Simplification of the coupling term}
\label{apdx:simpli}
In this appendix we simplify the expressions obtained in appendix 
\ref{apdx:vecppe} to get an estimate of the coupling coefficient, \ie 
of the efficiency of non-linear coupling.
First we estimate  the constants $\beta$, $\eta^2$, and $\Sigma_2$ which 
appear in equation (\ref{eqn:coupldetail}).
$\beta$ can be written as 
$I_3/\Sigma$ (see appendix \ref{apdx:vecppe}).
From the hydrostatic equilibrium of the unperturbed state we find:
 \[4\pi G\int_0^z\rho_0(z')dz'+a^2\frac{\partial_z\rho_0}{\rho_0}
 =\int_0^z 4\pi G\rho_0 - \partial^2_{z_2}\phi_0\,dz'=-\mu^2z\]
This gives from the definition of \(I_3\):
\[I_3 = \int_{-\infty}^{+\infty}\rho_0\mu^4z^2dz \simeq \mu^4\Sigma H^2\]
and thus:
\[\beta \simeq \mu^4 H^2\]
Now we have to compute $4\pi G\Sigma_2/\Sigma$.
One can say that $\Sigma_2/\Sigma$ is of the order of magnitude of
$\rho_m$, which is the density in the mid-plane of the disk.
Now $4\pi G\rho_m$ is the frequency of vertical oscillations of a test
particle in the rest potential of the galactic disk.
We note this frequency $\nu_z$, where the $z$ index reminds 
that this frequency depends on the vertical excursion of the test particle.
Hence:

\[\eta^2 \sim \mu^2 - \nu_z^2\]

Finally we consider $\Theta = \kappa\tomega/(\tomega^2-\kappa^2)$. 
$\Theta$ is of the order of 1, except in the vicinity of
the Lindblad resonances, where it become large. This is essentially the 
same effect that was found in previous works (Tagger \etal, 1987; Sygnet 
\etal, 1988) to make non-linear coupling very efficient near the Lindblad 
resonances of the coupled waves (although in that case the coupling 
coefficient was obtained in a kinetic rather than fluid description, \ie 
adapted to the stellar rather than gaseous component of the disk). 
On the other hand, our assumption of 
weak non-linearities certainly breaks down if $\Theta$ becomes too large, 
\ie exactly at the Lindblad resonances. A convenient estimate is that this break-down 
occurs at one sound-crossing time $a/\kappa$ (or equivalently one 
epicyclic radius) from the resonances, giving an upper limit for $\Theta$:
\[\Theta_{\mbox{\scriptsize max}} \sim \frac{\kappa^2}{a^2q^2} \sim
\frac{1}{q^2H^2}
\sim 10 \mbox{ to } 100\]

Now we can estimate the different terms of equation
{\ref{eqn:coupldetail}. The first term can be estimated as follows:

\begin{itemize}
\item{
The term in inner brackets is of the order of $2$. This estimate  is of course
expected to vary of a factor $2$ or $3$, but no more. In particular,
it is irrelevant to speculate on a possible resonant effect due to a 
vanishing value of
$q_S$: as we will find below, the spiral is tightly wound in the 
region of strong coupling.}
\item{It is an easy matter to see that $\Theta_1\Theta_2$ is always about
$0.5$, even in the vicinity of the Lindblad resonances, since
$\tomega_1+\tomega_2 = \tomega_S$.}
\item{Finally, $\beta$ is of the order of $\mu^4H^2$, as seen above.}
\end{itemize}

One can then deduce the order
of magnitude of the first term ${\cal T}_1$:

\begin{equation}
\label{eqn:t1}
{\cal T}_1 = \mu^4H^2\frac{q_1q_2}{\kappa^2}\tomega_S
\end{equation}

Let us evaluate now the second term ${\cal T}_2$:

\begin{itemize}
\item{First, we compare $\nu_z^2$ and $\mu^2$. 
We have:

\[\frac{\nu_z^2}{\Omega^2} \simeq \frac{4\pi G\rho_m}{GM/r^3} \simeq \frac{4r}{H} \sim 40\]

Now $\Omega$ is larger than $\mu$ in the region of the galaxy where the
rotation curve is nearly flat, hence 
$\nu_z^2 \gg \mu^2$.}

\item{On the other hand, since $\tomega\sim\Omega$,  we have $\nu_z^2 \gg \tomega_1\tomega_2$.}

\item{Finally, we assume that $q_2 < q_S$ (this is a good approximation in the region of
strong coupling; we will emphasize it again when we study the 
localization of the coupling
).} 
\end{itemize}

We deduce that:

\begin{equation}
\label{eqn:t2}
{\cal T}_2 = \nu_z^2\tomega_S
\end{equation}

Let us note that ${\cal T}_1 / {\cal T}_2$, from eqs. (\ref{eqn:t1}) and (\ref{eqn:t2}),
is of the order of:

\[\frac{\mu^2}{\nu_z^2}\frac{\mu^2}{\kappa^2}(q_1H)(q_2H)\]

where each factor is smaller or much smaller than unity. Hence the first term
is always negligible compared to the second one.

Let us now find an estimate for ${\cal T}_3$. We will derive it in the 
vicinity of a Lindblad resonance, maximizing $\Theta$. We note that:
\[\frac{\beta}{\mu^2a^2} = \frac{\mu^2}{\kappa^2} \ll 1\]
so that $\beta+a^2\mu^2 \simeq a^2\mu^2$.

We estimate:

\[\frac{{\cal T}_3}{{\cal T}_2} \simeq \frac{q_S}{\kappa}a^2\mu^2q_2\frac{1}{q_2^2H_2^2}
\frac{1}{\nu_z^2\tomega_S} \simeq  \underbrace {\frac{q_S}{q_2}}_{\sim 10}
\underbrace{\frac{\kappa}{\tomega_S}}_{\sim 1} \underbrace{\frac{\mu^2}{\nu_z^2}}_{\sim
\frac{1}{50}}\]

Hence  ${\cal
 T}_3$ is smaller or comparable with ${\cal T}_2$ 
in the vicinity of a Lindblad resonance. Away from the resonances, the ratio
${\cal T}_3/{\cal T}_2$ is still lower. Thus we deduce, since ${\cal T}_1$
is always negligible compared to ${\cal T}_2$, that the coupling term is always
of the order of ${\cal T}_2$:

\[\frac{d}{dt}E_1 \sim \nu_z^2\tomega_S Z_1Z_2\sigma\]

\section{Expression of energies}
\label{apdx:energy}

The coupling equation written in the previous appendix involves both the energy and
the amplitude of warp~1. They are actually linked by an expression we wish to derive.

The energy $E_1$ reads ({\it cf.} eq. (\ref{eqn:coupstep3})):

\[E_1 = \int_{-\infty}^{+\infty}\rho_0(z) (|U_1|^2+|V_1|^2+|W_1|^2 +a^2|s_1|^2+
\Re[s_1^*\phi_1])dz\]

After a straightforward calculation, in particular making use of the integrals
of appendix~\ref{apdx:vecppe}, one obtains:

\[E_1 = \biggl[\biggl(1+\frac{4\Omega^2}{\kappa^2}\biggr)q_1^2H^2\Theta_1^2
\frac{\mu^4}{\kappa^2} + \tomega_1^2 + \mu^2\biggr]\Sigma Z^2\]

Far from the Lindblad resonances, it is easy to see that the first term of
this expression of energy is negligible compared to the third, and
{\it a fortiori} to the second. However, at the Lindblad resonances of the
warp, the first term can become important and dominate the others.
The physical interpretation is that near the Lindblad resonance the 
kinetic energy associated with horizontal motions, due to the 
compressibility of the gas, becomes dominant.

Hence, at the Lindblad resonance, on can have energy ``hidden'' in the 
horizontal motions associated with the warp, \ie a large energy with a 
small vertical displacement. This is illustrated in figures
\ref{fig:verti} and \ref{fig:hori}.

\begin{figure}
\parbox{6.5cm}{
\psfig{file=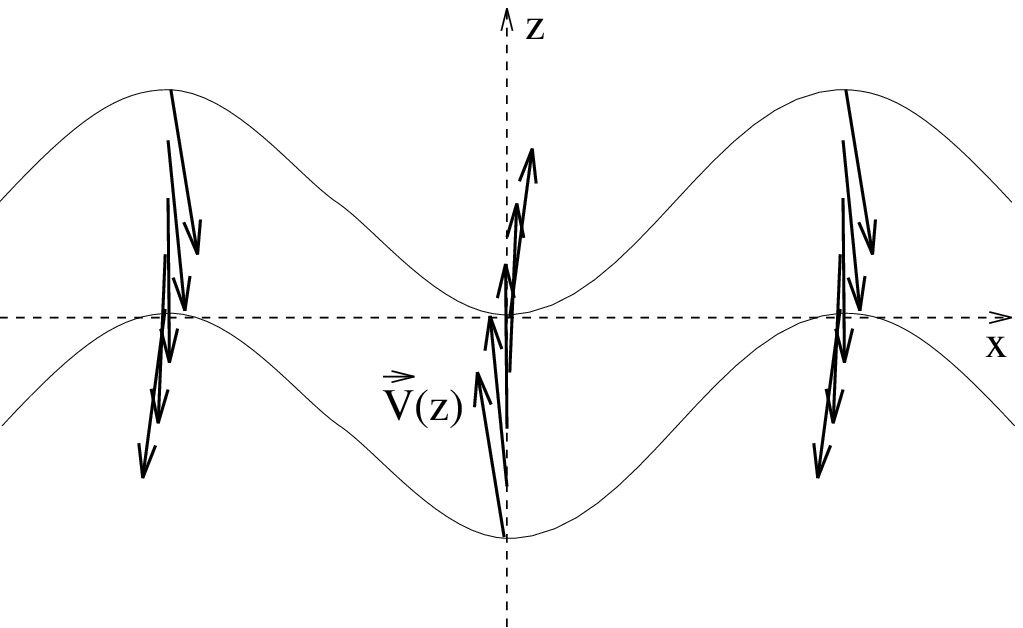,width=6.5cm}
\caption{\label{fig:verti}
\scriptsize
This figure shows the velocity field of a warp away from the Lindblad
resonance.}
}
\hfill
\parbox{6.5cm}{
\psfig{file=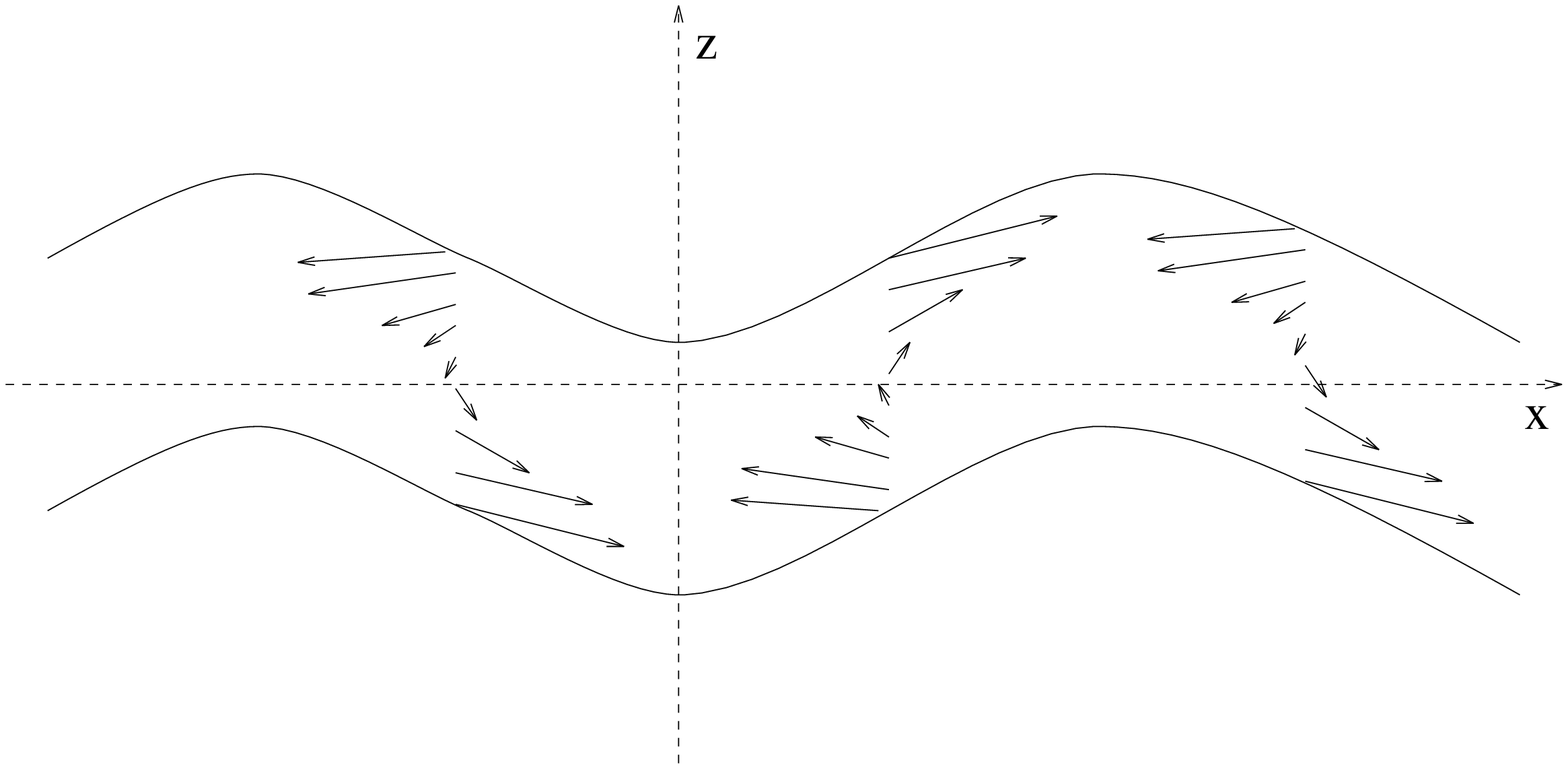,width=6.5cm}
\caption{\label{fig:hori}
\scriptsize
This figure shows the velocity field, which becomes nearly horizontal,
of a warp near a Lindblad resonance.}
}
\end{figure}

On the other hand, for a spiral, the vertical motion never dominates even 
when compressibility becomes important. This can be directly seen
from the expression of the energy of the spiral:

\[E_S = \biggl[\frac{\tomega_S^2}{q_S^2}\biggl(2+\frac{4\Omega^2}{\kappa^2}\biggr)-
\frac{\kappa^2}{q_S^2}\biggr]\frac{\sigma^2}{\Sigma}\]

which does not show any resonant term. Thus the perturbed surface density $\sigma$
represents fairly well the energy of the spiral wave. There are no hidden motions
(hidden in the sense that they don't have incidence on the observable)
similar to the hidden horizontal motions in warps. 
Thus we have:
\[K_i = \biggl[\biggl(1+\frac{4\Omega^2}{\kappa^2}q_i^2H^2\Theta_i^2\frac{\mu^4}{\kappa^2}
+\tomega_i^2+\mu^2\biggr]^{1/2} \mbox{ for i=1 or 2}\]
and
\[K_S = \biggl[\frac{\tomega_S^2}{q_S^2}\biggl(2+\frac{4\Omega^2}{\kappa^2}\biggr)
-\frac{\kappa^2}{q_S^2}\biggr]^{1/2}\]

\end{document}